\documentclass[nohyper,11pt,letterpaper]{JHEP3}
\usepackage[dvips]{epsfig}
\usepackage{amsmath}

\newcommand{\be}{\begin{equation}}
\newcommand{\ee}{\end{equation}}
\newcommand{\bea}{\begin{eqnarray}}
\newcommand{\eea}{\end{eqnarray}}
\newcommand{\ba}{\begin{eqnarray}}
\newcommand{\ea}{\end{eqnarray}}

\newcommand{\eqn}[1]{(\ref{#1})}
\newcommand{\beq}{\begin{equation}}
\newcommand{\eeq}{\end{equation}}
\newcommand{\beqa}{\begin{eqnarray}}
\newcommand{\eeqa}{\end{eqnarray}}
\newcommand{\beqar}{\begin{eqnarray*}}
\newcommand{\eeqar}{\end{eqnarray*}}

\newcommand{\eg}{{\it e.g.,}\ }
\newcommand{\ie}{{\it i.e.,}\ }

\newcommand{\sac}{\, , \qquad}

\newcommand{\wn}{\omega}
\def\nc {N_\mt{c}}
\def\nf {N_\mt{f}}

\def\t6 {T_\mt{D6}}

\def\gym {g_\mt{YM}}

\newcommand{\te}{x^0_\mt{E}}

\newcommand{\tf}{T_\mt{fun}}
\newcommand{\leff}{g_\mt{eff}}

\newcommand{\mq}{M_\mt{q}}      % Quark mass
\newcommand{\R}{L} % AdS radius
\newcommand{\mbar}{\bar{M}}

\newcommand{\mt}[1]{\textrm{\tiny #1}}

\newcommand{\uem}{U(1)_\mt{EM}}
\newcommand{\bk}{{\mathbf k}}
\newcommand{\jem}{J^\mt{EM}}
\newcommand{\A}{{\cal A}}

\newcommand{\tdec}{T_\mt{dec}}
\newcommand{\mmes}{M_\mt{mes}}
\newcommand{\aem}{\alpha_\mt{EM}}

\title{Bright branes for strongly coupled plasmas
}

\author{David Mateos  \\
Physics Department, University of California, Santa Barbara, CA 93106-9530, USA}

\author{Leonardo Pati\~no \\
Instituto de F\'\i sica, UNAM, Apdo. Postal. 20-364, 01000 D.F. M\'exico}

\abstract{We use holographic techniques to study photon production in a class of finite temperature, strongly coupled, large-$\nc$ $SU(\nc)$ quark-gluon plasmas with 
$\nf \ll \nc$ quark flavours. Our results are valid to leading order in the electromagnetic coupling constant but non-perturbatively in the $SU(\nc)$ interactions. The spectral function of electromagnetic currents and other related observables exhibit an interesting structure as a function of the photon frequency and the quark mass. We discuss possible implications for heavy ion collision experiments.}


\preprint{}

\begin{document}{\vskip 1cm}

%%%%%%%%%%%%%%%%%%%%%%%%%%%%%%%%%%%%%%%%%%%%%%%%%%%%%%%%%%%%%%
%%%%%%%%%%%%%%%%%%%%%%%%%%%%%%%%%%%%%%%%%%%%%%%%%%%%%%%%%%%%%%
%%%%%%%%%%%%%%%%%%%%%%%%%%%%%%%%%%%%%%%%%%%%%%%%%%%%%%%%%%%%%%
%%%%%%%%%%%%%%%%%% INTRODUCTION %%%%%%%%%%%%%%%%%%%%%%%%
%%%%%%%%%%%%%%%%%%%%%%%%%%%%%%%%%%%%%%%%%%%%%%%%%%%%%%%%%%%%%%
%%%%%%%%%%%%%%%%%%%%%%%%%%%%%%%%%%%%%%%%%%%%%%%%%%%%%%%%%%%%%%
%%%%%%%%%%%%%%%%%%%%%%%%%%%%%%%%%%%%%%%%%%%%%%%%%%%%%%%%%%%%%%
\section{Introduction}
Photons emitted by the quark-gluon plasma (QGP) created at the Relativistic Heavy Ion Collider (RHIC), as well as in future heavy ion experiments at the Large Hadron Collider (LHC), are expected to be an important probe of the plasma physics. Because of the finite extent of the plasma ball and of the weakness of the electromagnetic coupling constant, once a photon is produced it escapes essentially undisturbed, thus carrying valuable information about local properties of the plasma at the point of emission 
\cite{stankus}.

At weak coupling, photon production in Quantum Chromodynamics (QCD) can be studied by means of perturbation theory \cite{perturbative}. However, results from RHIC seemingly indicate that the plasma created there does not behave as an almost free gas of quarks and gluons, but rather as a strongly coupled liquid 
\cite{shuryak}. Consequently, understanding the spectrum and rate of photon emission requires a non-perturbative calculation. This is currently beyond the scope of analytical methods, and performing this calculation on the lattice is problematic due to the real-time, Lorentzian-signature nature of the physics involved.

In view of the above, it is interesting to investigate photon production in the context of the 
gauge/gravity correspondence \cite{malda}. Although the gravity dual of QCD itself is not known, the duals of a large class of finite-temperature, strongly coupled,  large-$\nc$ $SU(\nc)$ gauge theories are known. One may hope that generic, universal features exhibited by such theories may be useful indications as to the behaviour of QCD, at least in the regime of strong coupling. With this in mind, ref.~\cite{CKMSY} studied photon production at strong coupling in ${\cal N}=4$ super Yang-Mills (SYM), a four-dimensional theory with massless matter in the adjoint representation, and ref.~\cite{PS} calculated the rate of photon emission in a four-dimensional theory with massless quarks. One purpose of this paper is to study the effect of a non-zero quark mass on photon production.

In perturbative, weak-coupling calculations it is usually assumed that the temperature $T$ is high enough so that the thermal quark mass, of order $\sqrt{\lambda} T$ (with 
$\lambda = \gym^2 \nc$ the 't Hooft coupling) is much larger than the bare, zero-temperature quark mass $\mq$, and so the latter is effectively set to zero. Regardless of the precise temperature range in which this is applicable, the very nature of this approximation assumes that quarks are well defined quasi-particles to which physical properties such as a thermal mass can be meaningfully assigned. 
This is true at sufficiently weak values of the coupling, but it is not necessarily true at strong coupling. Well defined quasi-particle excitations certainly do not seem to exist in the phases of interest of the holographic theories that we will consider \cite{prl,thermo}, and they might not exist either in a QCD plasma at temperatures just above deconfinement. Under these circumstances, the bare quark mass is better thought of as a microscopic parameter on which the physics depends, but without a direct interpretation as the mass of a physical quasi-particle. Examples of physical quantities that depend on the quark mass through the dimensionless combination $\mq/T$ include the entropy density \cite{prl,thermo}, the shear viscosity \cite{viscosity}, etc. 
We will see that the same is true for observables related to photon production.

We will consider finite-temperature, $SU(\nc)$ super Yang-Mills (SYM) theories coupled to $\nf$ flavours of fundamental matter whose bare mass is an adjustable parameter. The fundamental matter includes both fermions and scalars, which we will refer to collectively as `quarks'. Our goal is to study photon production as a function of the quark mass and the photon frequency, to leading order in the electromagnetic coupling constant but non-perturbatively in the $SU(\nc)$ interactions. Note that, in addition to the quarks, the matter content in these theories also includes adjoint fields. Since we would like to model as closely as possible QCD, which has no adjoint matter, we will assume that the photon couples directly only to the quarks, \ie we will assign vanishing electric charge to the adjoint matter.

In the limit of a small number of flavours, $\nf \ll \nc$, the theories above possess a simple dual description in terms of $\nf$ Dq-brane probes in the gravitational background of $\nc$ Dp-branes \cite{flavour}. At temperatures high enough so that the gluons (and the adjoint matter) are deconfined, the gravitational background contains a black hole \cite{witten}, represented by the shaded blob in fig.~\ref{embeddings}. In this deconfined phase, the fundamental matter may still be in two different phases separated by a first-order phase transition \cite{thermo}.\footnote{Specific examples of this transition were originally seen in \cite{johanna,us}. Aspects of these transitions were independently studied in the D3/D7 system in \cite{recent,recent2}, and in a slightly different framework in \cite{recent8}.}
\FIGURE{
\includegraphics[width=0.95 \textwidth]{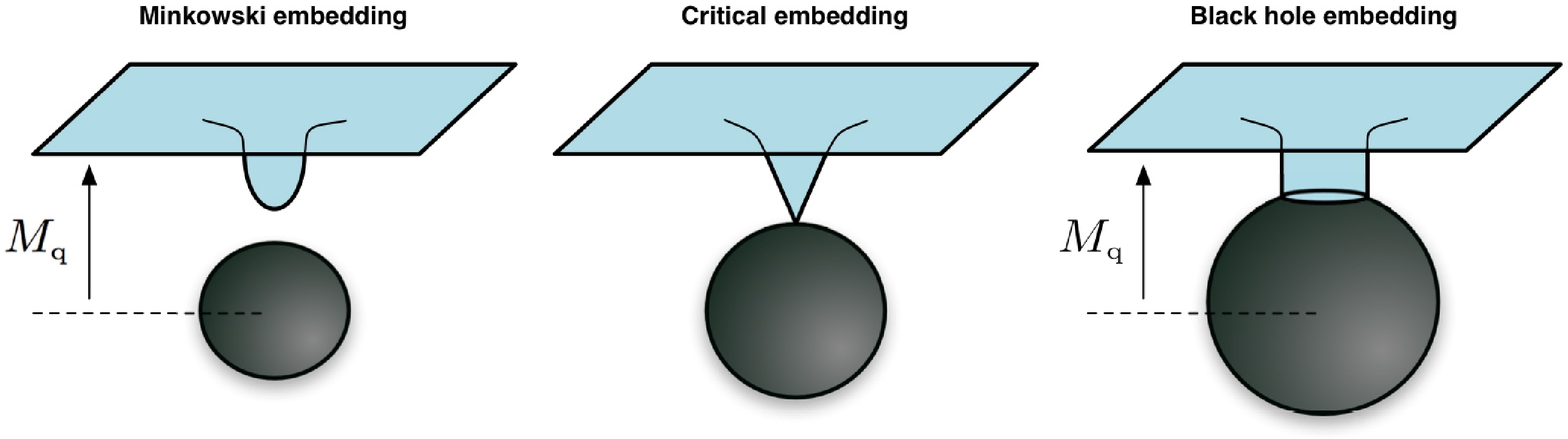}
\caption{Possible embeddings for Dq-brane probes in the background of black Dp-branes. The critical embedding is skipped over by the phase transition.}
\label{embeddings}}

From the viewpoint of the holographic description, the basic physics
behind this transition is easily understood. The asymptotic distance between the 
Dq-branes and the black hole is proportional to the quark mass, whereas the 
size of the black hole horizon is proportional to the temperature. 
Thus, for sufficiently small $T/\mq$, the Dq-branes are deformed by the gravitational attraction of the black hole, but remain entirely outside the horizon in what we call a `Minkowski embedding' (see fig.~\ref{embeddings}). However, above a critical temperature $\tf$, the gravitational force overcomes the tension of the branes and these are pulled into the horzion. We refer to such configurations as `black hole embeddings'.
In between these two types of embeddings there exists a `critical embedding' in which the branes just `touch the horizon at a point'. However, thermodynamic considerations reveal that a first-order phase transition occurs between a Minkowski and a black hole embedding. In other words, the critical embedding is skipped over by the phase transition, and near-critical embeddings turn out to be metastable or unstable. 

In the dual field theory, this phase transition is exemplified by discontinuities in physical quantities such as, for example, the quark condensate or the contribution of the fundamental matter to the entropy density. However, the most striking feature of this phase transition is found in the spectrum of physical excitations of the fundamental matter. 
In the low-temperature, Minkowski phase the spectrum is gapped and contains a discrete set of deeply bound mesons (\ie quark-antiquark bound states) with masses of order 
$\mmes \sim \mq / \sqrt{\lambda}$. These mesons are dual to excitations supported on the probe branes -- see, \eg \cite{us-meson,holomeson,holomeson2}. In addition, the Minkowski-phase spectrum also contains well defined, 
quark-like excitations described by strings stretching between the tip of the branes and the horizon. These have masses of order $\mq$ and are therefore parametrically heavier than the mesons. Both sets of excitations are absolutely stable in the large-$\nc$, strong coupling limit under consideration. 

In the high-temperature, black hole phase stable mesons cease to exist. Rather one finds a continuous and gapless spectrum of excitations \cite{prl,thermo,melt,spectre}. Hence the mesons dissociate or `melt' at the first order phase transition at 
$T_\mt{fun}$, and beyond  it no well defined, quasi-particle notion of an individual quark exists, since a string stretching between any point on the branes and the horizon will quickly fall through the horizon. In the gauge theory this corresponds to the fact that any localised quark charge will quickly spread across the entire plasma, thus loosing its identity.
 
In this paper we will study photon production in the black hole phase. We will see that this is proportional to $\nf \nc$, \ie to the number of electrically charged degrees of freedom, as one would expect. For $\mq=0$ our result coincides with that of \cite{CKMSY}, except for the fact that in that reference the overall normalisation is set by $\nc^2$, reflecting the fact that electric charge is carried by adjoint degrees of freedom. We stress that the photon production we will calculate is not a correction to an $O(\nc^2)$-result but the leading result in the large-$\nc$ expansion, since in our case the adjoint matter is electrically neutral. We will see that the gravity picture provides a simple, geometric explanation for the agreement (up to normalisation) between our $\mq=0$ result and that of \cite{CKMSY}. 

We will not explicitly study photon production in the Minkowski phase, since we have a clear expectation of what the result would be. Indeed, as explained above, in this phase 
the spectrum consists of a discrete set of stable modes, so we expect the spectral function \eqn{spectral} to be given by a sum of delta functions supported at appropriate values of the momentum. We will confirm this expectation indirectly by calculating the spectral function for black hole embeddings near the critical embedding (see fig.~\ref{embeddings}). 

In order to study photon production, one must in principle introduce a dynamical photon that couples appropriately to the $SU(\nc)$ fundamental matter and construct the dual of the $SU(\nc) \times \uem$ gauge theory thus obtained. However, we will see that, to leading order in the electromagnetic coupling constant, photon production in the $SU(\nc) \times \uem$ theory is completely determined by correlators in the original $SU(\nc)$ theory with no dynamical photon. This observation allows the holographic calculation of the desired correlators.  

This paper is organised as follows. In sec.~\ref{hot} we discuss how to calculate the rate of photon production in a gauge theory at finite temperature. In sec.~\ref{holo} we explain how to use the holographic description in terms of a Dp/Dq system to study photon production in the dual gauge theory. In secs.~\ref{d3d7} and \ref{d4d6} we present our results for the D3/D7 and the D4/D6 systems, respectively. We conclude in sec.~\ref{discussion} with a discussion of our results and their possible implications for heavy ion collision experiments.

\section{Photon emission in a hot gauge theory}
\label{hot}
Consider a 
$(d+1)$-dimensional $SU(\nc)$ gauge theory with $\nf$ flavours of fundamental fermions $\Psi^a$ and scalars $\Phi^a$, $a=1, \ldots, \nf$, to which we will collectively refer as `quarks'. Imagine coupling this theory to electromagnetism by introducing and additional, dynamical, Abelian gauge field, the photon, that couples to the fundamental fields with strength (electric charge) $e$.\footnote{For simplicity we assume that all the fundamental fields have equal mass and carry equal electric charge, but our results can be generalised straightforwardly. Later we will consider $SU(\nc)$ theories that also contain adjoint matter. Since our goal is to model QCD, which has no such matter, we will assign vanishing electric charge to the adjoint fields.} 
This new theory may be constructed by adding a kinetic term for the photon and replacing the $SU(\nc)$-covariant derivative $D_\mu$ by  
${\cal D}_\mu = D_\mu - i e \A_\mu$ when acting on the fundamental fields, with 
$\mu=0, \ldots, d$. In this way we obtain an $SU(\nc) \times \uem$ gauge theory with Lagrangean
\be
{\cal L} = {\cal L}_{SU(\nc)} - \frac{1}{4} {\cal F}_{\mu \nu}^2 + e \A^\mu J^\mt{EM}_\mu \,,
\ee
where ${\cal F}_{\mu\nu} = \partial_\mu \A_\nu - \partial_\nu \A_\mu$ and the electromagnetic current is given by 
\be
\jem_\mu = \bar{\Psi} \gamma_\mu \Psi 
+ \frac{i}{2} \Phi^* \left( {\cal D}_\mu \Phi \right) 
- \frac{i}{2} \left( {\cal D}_\mu \Phi \right)^* \Phi \,. 
\label{current}
\ee
A sum over flavour and colour indices is implicit in this formula. Note that the photon field enters this current through the covariant derivative acting on the scalars. 

In thermal equilibrium, the differential photon emission rate per unit time and volume, at leading order in the electromagnetic coupling constant $e$, is then given by \cite{bellac}
\be
\frac{d \Gamma}{d^d {\bk}} =  \frac{e^2}{(2\pi)^d \, 2 |{\bk}|} \,  n_\mt{B} (k^0) \,  
\sum_{s=1}^{d-1} \epsilon^\mu_{(s)} ({\bk}) \epsilon^\nu_{(s)} ({\bk}) 
\left. \chi_{\mu \nu} (k)  \right|_{k^0 = \bk}
\,, \label{dgamma}
\ee
where $k = (k^0, \bk)$ is the photon null momentum,
\be
\chi_{\mu\nu} (k) = -2 \, \mbox{Im} \, G^\mt{R}_{\mu\nu} (k)
\label{spectral}
\ee
is the spectral density, and
\be
G^\mt{R}_{\mu \nu} (k) = - i \int d^{d+1} x \, e^{-i k \cdot x} \, \Theta (x^0) 
\langle [ \jem_\mu (x), \jem_\nu (0) ] \rangle 
\label{green}
\ee
is the retarded correlator of two electromagnetic currents, whose diagramatic representation (including photon fields as external legs) is given in fig.~\ref{loops}. 
\FIGURE[t!]{
\begin{tabular}{cc}
\includegraphics[width=0.63 \textwidth]{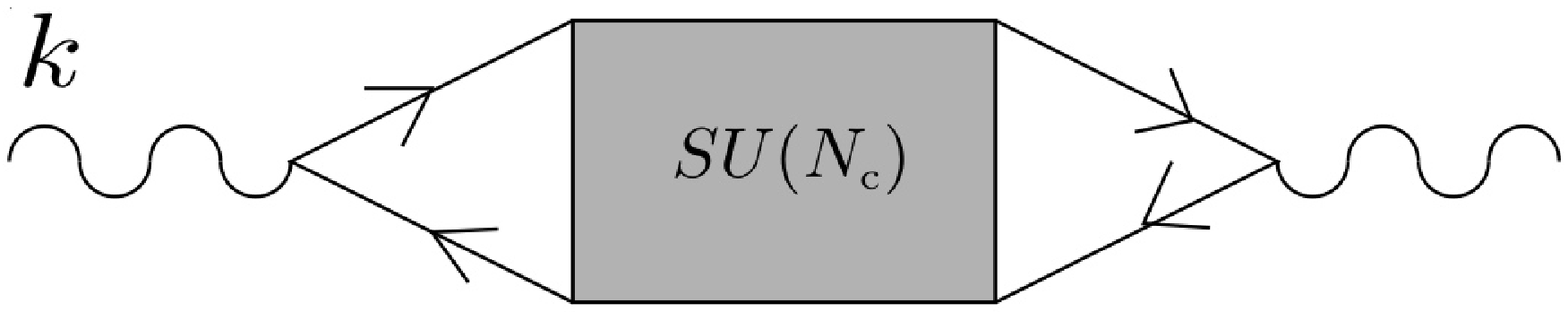} &  
\includegraphics[width=0.37 \textwidth]{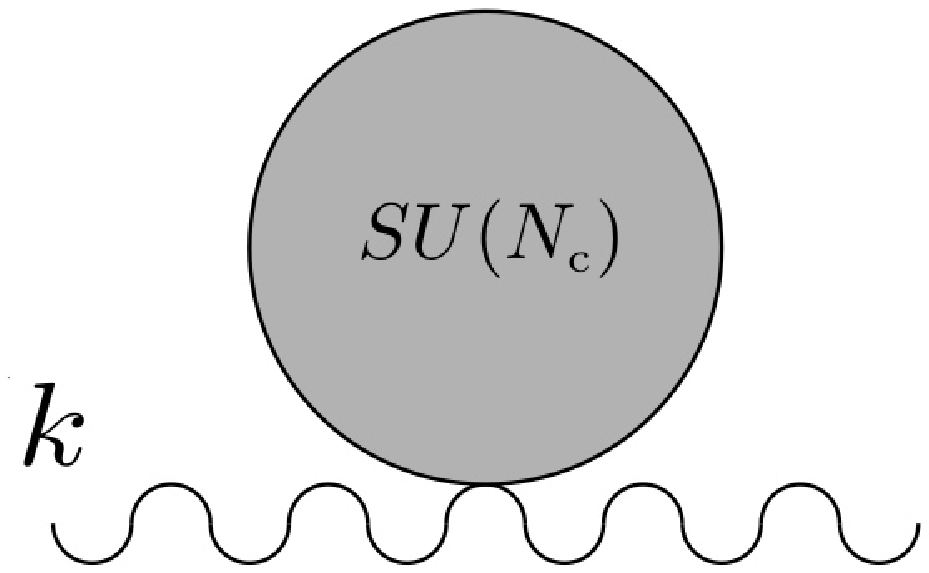} \\
 (a) & (b)\\
\end{tabular}
\caption{Diagrams contributing to the two-point function of electromagnetic currents  
\eqn{green}. The external line corresponds to a photon of momentum $k$. As explained in the text, to leading order in the electromagnetic coupling constant only $SU(\nc)$ fields `run' in the loops represented by the shaded blobs.}
\label{loops}}
Finally,
\be
n_\mt{B} (k^0) = \frac{1}{e^{k^0/T} -1}
\ee
is the standard Bose-Einstein distribution function. Without loss of generality we will assume that $\bk$ points in the $x^1$-direction, and we will denote by 
$x^i$, $i=2, \ldots , d$ the remaining spatial directions. The polarisation vectors  
$\epsilon^\mu_{(s)}$ may be chosen to be unit spatial vectors orthogonal to ${\bf k}$. Transversality of the correlator 
implies that the sum over polarisation vectors in \eqn{dgamma} may be replaced by the trace of the spectral function, \ie
\be
\chi^\mu_{\,\, \mu} (k) \equiv \eta^{\mu\nu} \chi_{\mu\nu} (k) =  
\sum_{s=1}^{d-1} \epsilon^\mu_{(s)} ({\bk}) \epsilon^\nu_{(s)} ({\bk}) \,
\chi_{\mu \nu} (k) \,,
\ee
and rotational invariance ensures that this is given by 
\be
\chi^\mu_{\,\, \mu} (k) = -2 (d-1) \, \mbox{Im} \, G^\mt{R} (k) \,,
\label{trace}
\ee
where 
\be
G^\mt{R} (k) \equiv \frac{1}{d-1} \, \delta^{ij} G^\mt{R}_{ij} (k) \,.
\label{G}
\ee
The trace of the spectral function also determines the electric conductivity as
\be
\sigma = \frac{e^2}{2(d-1)} \lim_{k^0 \rightarrow 0} 
\frac{1}{k^0} \left. \chi^\mu_{\,\, \mu} (k)\right|_{k^0 = \bk} \,.
\label{sigma}
\ee

Thus in order to study photon production we must in principle calculate the two-point function \eqn{green} in the $SU(\nc) \times \uem$ theory. However, as noted in \cite{CKMSY}, to leading order in the electromagnetic coupling constant this reduces to a calculation purely within the original $SU(\nc)$ theory; this is the key observation that will allow us to perform this calculation using the dual gravitational description, since the dual of the $SU(\nc) \times \uem$ theory is unknown. To see this, note first that the terms in the electromagnetic current \eqn{current} proportional to the photon field (implicit in the covariant derivative) lead to higher-order contributions in $e$ to the correlator
\eqn{green}, and can thus be ignored to leading order in $e$. Second, observe that the two-point function of the remaining terms in the current can be calculated in the $SU(\nc)$ theory, since again the effects of including the dynamical photon are of higher order in $e$. Diagramatically, this means that no photon fields are present in the shaded blobs in fig.~\ref{loops}. An additional observation is the fact that the tadpole diagram in fig.~\ref{loops} has no imaginary part and hence does not contribute to the desired spectral function.

We therefore conclude that, to leading order in the electromagnetic coupling constant,  photon production in an $SU(\nc) \times \uem$ theory is completely determined by the two-point function of the electromagnetic current in the $SU(\nc)$ theory. In the rest of the paper we will calculate this correlator in $SU(\nc)$ SYM theories coupled to fundamental matter.

\section{Holographic description}
\label{holo}

In the decoupling limit, the black Dp-brane solution in the string frame takes the form 
\bea
ds^2 &=& H^{-1/2} \left( -f dx_0^2 +  d{\bf x}^2 \right) + H^{1/2} \left( \frac{dr^2}{f}
+ r^2 d\Omega_{\it 8-p}^2 \right) \label{metric} 
\,, \nonumber\\
e^\Phi &=&  H^{\frac{3-p}{4}} \sac\qquad  C_{01\ldots p} = H^{-1} \,, 
\eea
where ${\bf x} = (x^1,\ldots, x^p)$, 
\be
H(r) = \left( \frac{L}{r} \right)^{7-p} \sac f(r) =1- \left( \frac{r_0}{r} \right)^{7-p} \,,
\ee
and $L$ is a length scale (the AdS radius in the case $p=3$). The horizon lies at $r=r_0$. 
As usual, regularity of the Euclidean section, obtained through $x^0 \rightarrow i \te$, requires
that $\te$ be identified with period
\be 
\frac{1}{T} = \frac{4\pi  \R}{7-p} \left( \frac{\R}{r_0}
\right)^{\frac{5-p}{2}} \,. 
\label{beta} 
\ee
In this paper it will be useful to work with a different radial coordinate $u$ related to $r$ through 
\be
u=\frac{1}{2} \left( \frac{r_0}{r} \right)^{\frac{7-p}{2}} \,,
\label{u}
\ee
in terms of which $f=1-4 u^2$ and the horizon lies at $u=1/2$.

According to the gauge/gravity correspondence, string theory on the
background above is dual to $(p+1)$-dimensional, maximally supersymmetric
Yang-Mills theory at temperature $T$. In some cases one periodically
identifies some of the `Poincare' directions ${\bf x}$ in order to
render the theory effectively lower-dimensional at low energies; a
prototypical example is that of a D4-brane with one compact space
direction. Under these circumstances a different background with no
black hole may describe the low-temperature physics, and a phase
transition may occur as $T$ increases \cite{witten}. In the gauge
theory this is typically a confinement/deconfinement phase
transition for the gluonic (and adjoint) degrees of freedom.
Throughout this paper we assume that $T$ is high enough, in which
case the gauge theory is in the deconfined phase and the appropriate 
gravitational background is \eqn{metric}.

Consider now $\nf \ll \nc$ coincident Dq-brane probes that share 
$d$ spacelike Poincar\'e directions with the background Dp-branes 
and wrap an $S^n$ inside the $S^{8-p}$. 
We will assume that the Dq-branes also extend along
the radial direction, so that $q=d+n+1$. We will denote by 
$x^\mu$, with $\mu=0, \ldots, d$, the directions common to both branes.
In the gauge theory the Dq-branes correspond to introducing $\nf$ 
flavours of fundamental matter of equal mass that propagate along
a $(d+1)$-dimensional defect. To ensure stability, we will assume
that the Dp/Dq intersection under consideration is supersymmetric at
zero temperature. Under these conditions the Ramond-Ramond field sourced by the
Dp-branes does not couple to the Dq-branes. Two cases of special
interest here are the D3/D7  ($n=3$) \cite{johanna} and the D4/D6
($n=2$) \cite{us} systems. If one of the D4 directions is compact,
then both cases can effectively be thought of as describing the
dynamics of a four-dimensional gauge theory with fundamental matter.

The $U(\nf)$ gauge symmetry on the Dq-branes is a global, flavour symmetry of the gauge theory. If this theory is coupled to electromagnetism as explained in the previous section (in particular, if all flavours are assigned equal electric charge $e$), then the electromagnetic current  $\jem_\mu$ is the conserved current associated to the overall $U(1) \subset U(\nf)$. In order to study photon emission to leading order in $e$ it suffices to evaluate the correlator \eqn{green} in the $SU(\nc)$ gauge theory with no dynamical photon.

At strong 't Hooft coupling and large $\nc$, this correlator is easily calculated holographically. Indeed, global symmetries of the gauge theory are in one-to-one correspondence with gauge symmetries on the gravity side, and each conserved current of the gauge theory is dual to a gauge field on the gravity side 
\cite{witten-holo}. Let $A_m$, $m=0, \ldots, q$, be the guage field associated to the overall $U(1)$ gauge symmetry on the Dq-branes. Upon dimensional reduction on the $n$-sphere wrapped by the Dq-branes, $A_m$ gives rise to a massless gauge field 
$\{ A_\mu, A_u \}$, $n$ massless scalars, and a tower of massive Kaluza-Klein 
(KK) modes. All these fields propagate on the $d+2$ non-compact dimensions of the Dq-branes. We will work in the gauge $A_u=0$, and we will consistently set to zero the scalars and the higher KK modes, since they are not of interest here. The gauge field 
$A_\mu$ is the desired dual to the conserved electromagnetic current $\jem_\mu$ of the gauge theory; just like the quarks are localised on a defect inside the gauge theory, the dual gauge field is localised on the Dq-branes inside the bulk spacetime. 

According to the gauge/gravity correspondence, correlation functions of $\jem_\mu$ can be calculated by varying the string partition function with respect to the value of $A_\mu$ at the boundary of the spacetime  
\eqn{metric}. Under the present circumstances, the string partition function reduces to $e^{iS}$, where $S$ is the sum of the supergravity action and the effective action for the Dq-branes. Since $A_\mu$ does not enter the supergravity action, the form of this action will not be needed here. Moreover, the Wess-Zumino part of the Dq-branes action does not contribute for the brane orientations and the gauge field polarisations considered in this paper. Therefore, it suffices to consider the Dirac-Born-Infeld part of the Dq-branes action:
\be
S = - \nf T_\mt{Dq} \int_{Dq} d^{q+1}x \, e^{-\phi} \sqrt{-\det (g+2\pi \ell_s^2 F)} \,,
\label{action}
\ee
where $g$ is the induced metric on the Dq-branes, $F=dA$ is the overall $U(1)$ field strength, $T_\mt{Dq} = 1/(2\pi \ell_s)^q g_s \ell_s$ is the Dq-brane tension, and $\ell_s$ and $g_s$ are the string length and coupling constant, respectively. Since we are only interested in the two-point function \eqn{green}, we only need to keep terms up to quadratic order in the gauge field. We will therefore work with the simpler action
\be
S = - \nf T_\mt{Dq} \int_{Dq} d^{q+1}x \, e^{-\phi} \sqrt{-\det g} 
\left( 1 + \frac{(2\pi \ell_s^2)^2}{4} F^2 \right) \,,
\label{sdq}
\ee
where $F^2 = F_{mn} F^{mn}$. 

The Dq-branes wrap an $n$-sphere in the directions transverse to the Dp-branes, 
so it is convenient to write the metric on $S^{8-p}$ in adapted coordinates as 
\be 
d\Omega_{\it 8-p}^2 = d\theta^2 + \sin^2 \theta \, d\Omega_{\it n}^2 
+ \cos^2 \theta \, d\Omega_{\it 7-p-n}^2  \,.
\label{sphere8}
\ee
Setting $\psi = \cos \theta$, the Dq-branes embedding may then be specified as $\psi=\psi(u)$. 
Asymptotically, \ie as $r \rightarrow \infty$ or $u \rightarrow 0$, this behaves as 
\cite{prl,thermo}
\be
\psi(u) \simeq m \, u^\frac{2}{7-p} + c \, u^\frac{2n}{7-p} + \cdots \,,
\ee
where $m$ and $c$ are proportional to the quark mass and condensate, respectively  (see below). 

Since the gauge field enters the action \eqn{sdq} quadratically, turning on $A$ induces a correction $\delta \psi$ of order $A^2$ on the branes embedding. When the action 
is evaluated on-shell, this generates a boundary-term contribution of the form 
\be
\delta S = \left. \frac{\partial {\cal L}}{\partial \psi'} \, \delta\psi \right|_\mt{boundary} \,,
\ee
where ${\cal L}$ is the Dq-branes Lagrangean density and $\psi'=d\psi/du$. 
Since $\delta \psi = {\cal O}(A^2)$, this has in principle the right form to contribute to the two-point function \eqn{green}. However, as explained in \cite{thermo}, this contribution is proportional to $\delta m$, and hence vanishes because the quark mass must be kept fixed in computing the correlator. 

We thus conclude that, in order to calculate the photon emission rate, we may consistently proceed by first determining the Dq-branes embedding in the absence of the gauge field, and then solving for the gauge field on that embedding. As explained above, we will set to zero the components of the gauge field on the $n$-sphere wrapped by the Dq-branes. Moreover, following \cite{PSS}, we will choose the gauge $A_u=0$, Fourier-decompose the $d+1$ remaining components of the gauge field as
\be
A_\mu(x^0, {\bf x}, u) = \int \frac{d k^0 d^d \bk}{(2\pi)^{d+1}} \, 
e^{-i k^0 x^0 + i \bk \cdot {\bf x}} \, A_\mu (k^0, \bk, u) \,, 
\label{fourier}
\ee
and choose $\bk$ to point in the $x^1$-direction. Under these circumstances the equations of motion for the $A_i$-components of the gauge field $(i=2,\ldots, d)$ that follow from the action \eqn{sdq} decouple from each other and from those for $A_0, A_1$. Setting 
$A_i \equiv A (k,u)$, the equation of motion takes the form
\be
\partial_u \Big[ Q(u) \, \partial_u A \Big] + k_0^2 \Big[1-f(u) \Big] P(u) A =0 \,,
\label{eom}
\ee 
where we have made use of the fact that $\left| {\bf k} \right|=k^0$. The functions 
$P(u), Q(u)$ depend on the embedding $\psi(u)$ and will be specified below. 
Eq.~\eqn{eom} may be consistently derived from the `reduced' action
\be
S =  - \tilde{\cal N}_\mt{Dq} \int dx^0 \, d^d {\bf x} \, du 
\left[ - P(u) ( \partial_0 A)^2 + f P(u) (\partial_1 A)^2  + Q(u) (\partial_u A)^2 \right] \,,
\label{d7s}
\ee 
obtained from \eqn{sdq} by integrating over the $n$-sphere and setting $\partial_i A =0$.
We will see below that the normalisation constant in front of this action scales as 
$\tilde{\cal N}_\mt{Dq} \sim \nf \nc T^{d-1}$ and determines the overall magnitude of photon production. 

According to the prescription in \cite{SS,HS}, the retarded correlator \eqn{G} is then given by 
\be
G^\mt{R}(\omega) = -  \tilde{\cal N}_\mt{Dq}  \lim_{u \rightarrow 0}  
\frac{2 Q(u) A^*(\omega,u) \partial_u A(\omega, u)}{A^*(\omega, 0) A(\omega, 0)} \,,
\label{corre}
\ee
where $\omega \equiv k^0/2\pi T$ and $A(\omega,u)$ is a solution of the equation 
of motion \eqn{eom} obeying the incoming-wave boundary condition at the horizon. The imaginary part of $G^\mt{R}(\omega)$, which is all we need, is $u$-independent, and 
it is convenient to calculate it at the horizon instead of at the boundary:
\be
\mbox{Im} \, G^\mt{R}(\omega)  =  
- \frac{\tilde{\cal N}_\mt{Dq}}{\left| A(\omega, 0) \right|^2} \, 
\mbox{Im} \lim_{u \rightarrow \frac{1}{2}}  
2 Q(u) A^*(\omega, u) \partial_u A(\omega, u) \,.
\label{imG}
\ee
Note that, although not explicitly indicated, several quantities above depend not only on the photon frequency $\omega$ but also on the quark mass, since the equation of motion \eqn{eom} that determines $A$ depends on the branes embedding through the functions $P$ and $Q$.

\section{The D3/D7 system}
\label{d3d7}

Here we will specialise the above discussion to the D3/D7 system.
This intersection is summarised by the array
\begin{equation}
\begin{array}{ccccccccccc}
   & 0 & 1 & 2 & 3 & 4& 5 & 6 & 7 & 8 & 9\\
\mbox{D3:} & \times & \times & \times & \times & & &  &  & & \\
\mbox{D7:} & \times & \times & \times & \times & \times  & \times & \times & \times &  &   \\
\end{array}
\label{D3D7}
\end{equation}
Of course, this is an interesting system because $d=3$, \ie both the gluons and
the fundamental fields in the gauge theory propagate in $3+1$ dimensions.

The metric \eqn{metric} for black D3-branes takes the form
\be
ds^2 = \frac{r^2}{L^2} \left( -f dx_0^2 +  d{\bf x}^2 \right) + 
\frac{L^2}{r^2} \frac{dr^2}{f} + L^2 d\Omega_{\it 5}^2 \,, 
\label{metricD3} 
\ee
where 
\be
L^4=4\pi g_s \nc \ell_s^4 \sac f=1- \frac{r_0^4}{r^4} \sac  r_0=\pi T L^2 \,. 
\ee
In this case the dimensionless coordinate \eqn{u} is\footnote{This is related to the coordinate used in \cite{PSS} as $2 u_\mt{here} = u_\mt{\cite{PSS}}$. \label{u-change}} $u=r_0^2/2 r^2$, in terms of which the metric becomes
\be
ds^2  = \frac{(\pi T L)^2}{2u} \left( -f dx_0^2 +  d{\bf x}^2 \right) + 
\frac{L^2}{4 u^2} \frac{du^2}{f} + L^2 d\Omega_{\it 5}^2 \,.
\ee
Specifying the D7-branes embedding through $\psi = \psi(u)$, as explained above, the induced metric takes the form
\be
ds^2_\mt{D7} = \frac{(\pi T L)^2}{2u} \left( -f dx_0^2 +  d{\bf x}^2 \right) + 
\frac{L^2 \left( 1- \psi^2 + 4 u^2 f \psi'^2 \right)}{4u^2 f (1-\psi^2)} \, du^2 +
L^2(1-\psi^2) d\Omega^2_{\it 3} \,. \label{g7}
\ee
In this case the asymptotic behaviour of $\psi$ is\footnote{In this limit, the coordinate $u$ used here and the coordinate $\rho$ used in \cite{prl,thermo} are related through 
$\rho^2 = 1/u$.} 
\be
\psi(u) = m \, u^{1/2} + c \, u^{3/2} + \cdots \,,
\label{asymp}
\ee
where the dimensionless constants $m$ and $c$ are related to the quark mass and condensate, respectively, through \cite{prl,thermo}:
\beqa 
\mq &=& \frac{r_0 m}{2^{3/2}\pi \ell_s^2} =
\frac{1}{2}\sqrt{\lambda}\,T\,m\,,\label{mqD3D7} \\
\langle {\cal O} \rangle &=& - 2^{3/2}\pi^3 \ell_s^2 \nf T_\mt{D7}
r_0^3\, c = -\frac{1}{8}\sqrt{\lambda}\,\nf\,\nc\,T^3\,c\,,
\label{cqD3D7} 
\eeqa
with $\lambda = \gym^2 \nc = 2\pi g_s \nc$ the 't Hooft
coupling. The operator ${\cal O}$ is a supersymmetric version of the quark
bilinear, and it takes the schematic form
\be {\cal O} = \bar{\Psi} \Psi + \Phi^\dagger X  \Phi + \mq \Phi^\dagger \Phi \,,
\label{oops} 
\ee
where $X$ is one of the adjoint scalars. We will loosely refer to
its expectation value as the `quark condensate'. A detailed
discussion of this operator, including a precise definition, can be
found in the appendix of ref.~\cite{density}.

Eq.~\eqn{mqD3D7} implies the relation $m=\bar{M}/T$ between
the dimensionless quantity $m$, the temperature $T$ and the mass
scale
\be 
\bar{M} = \frac{\sqrt{2} (2\pi \ell_s^2 \mq)}{\pi L^2} = \frac{2
\mq}{\sqrt{\lambda}}  = \frac{M_\mt{mes}}{2\pi} \,,
\label{mbarD3D7} 
\ee
where $\mmes$ is the mass of the lightest meson (\ie the mass gap) in the discrete
meson spectrum at zero temperature \cite{us-meson,holomeson,holomeson2,us}. 
$\mbar$ determines the scale of the temperature of the phase transition for the fundamental degrees of freedom, $\tf \sim \mbar$, since the latter takes place at 
$m \sim 1$.

The functions $P(u), Q(u)$ occurring in the reduced action \eqn{d7s} are given in this case by 
\be
P(u) = \frac{(1-\psi^2) \sqrt{1-\psi^2 + 4u^2 f \psi'^2}}{u f (2\pi T)^2}  \sac
Q(u) = \frac{(1-\psi^2)^2f}{2  \sqrt{1-\psi^2 + 4u^2 f \psi'^2}} \,,
\ee
and the normalisation constant is
\be
\tilde{\cal N}_\mt{D7} = {\cal N}_\mt{D7} \frac{4T (2\pi \ell_s^2)^2}{(\pi T L^2)^2} =
\frac{1}{4} \nf \nc T^2 \,, \label{normal}
\ee
where 
\be
{\cal N}_\mt{D7} = \frac{\nf T_\mt{D7} \Omega_{\it 3} r_0^4}{4T} = 
\frac{1}{32} \lambda \nf \nc T^3
\ee
is the normalisation constant introduced in \cite{prl,thermo} and 
$\Omega_{\it 3}=2\pi^2$ is the volume of a unit three-sphere. 

Analysis of eq.~\eqn{eom} near $u=1/2$ reveals that the solution obeying the incoming-wave boundary condition behaves as $A \sim (1-2u)^{-i\omega/2}$ near the horizon. We will thus seek a solution of the form
\be
A(\omega,u) = (1-2u)^{-i\omega/2} (1+2u)^{-\omega/2} \, F(\omega,u) \,,
\label{sol}
\ee
where $F$ is a regular function of $u$ and the second factor has been explicitly extracted for convenience. Since the differential equation solved by $A$ is linear, its overall normalisation is arbitrary, and it cancels out in eqs.~\eqn{corre}, \eqn{imG}. We choose to fix it by setting $F$ equal to unity at the horizon, \ie $F(\omega, \frac{1}{2})=1$. Using the form of the solution \eqn{sol} and this normalisation choice for $F$, the limit on the right-hand side of \eqn{imG} is readily calculated with the result
\be
\lim_{u \rightarrow \frac{1}{2}}  2Q(u) A^*(\omega, u) \partial_u A(\omega, u) = 
\frac{2 i \wn \left( 1- \psi_0^2 \right)^{3/2}}{2^\wn} \,,
\label{lim}
\ee
where $\psi_0 = \psi (1/2)$ is the value of $\psi$ at the horizon.
Thus the trace of the spectral density, eq.~\eqn{trace}, takes the form
\be
\chi^\mu_{\,\,\mu} (\omega) = -4 \, \mbox{Im} \, G^\mt{R}(\omega) = 
8 \tilde{\cal N}_\mt{D7}  \left( 1- \psi_0^2 \right)^{3/2} \,
\frac{\wn}{2^\wn \left| F(\omega,0) \right|^2} \,.
\label{trace1}
\ee

\subsection{Massless quarks}

Massless quarks correspond to the equatorial embedding $\psi=0$. In this case 
\be
P(u) = \frac{1}{u f(u) (2\pi T)^2} \sac Q(u) = \frac{1}{2} f(u) \,,
\ee
and the equation of motion \eqn{eom} becomes
\be
(1+2u)^2 (1-2u)^2 \, A'' - 8u (1+2u)(1-2u) \, A' + 8\wn^2 u \, A =0 \,,
\label{A}
\ee
where $\wn = k^0 / 2\pi T$ and a prime denotes differentiation with respect to $u$. 
Upon changing variables as described in footnote \ref{u-change}, the equation of motion above agrees precisely with eq.~(5.5d) in \cite{PSS}. The reason for this is that, for massless quarks, the induced metric \eqn{g7} on the D7-branes is exactly $AdS_5 \times S^3$. After reduction on the $S^3$, the quadratic action for the gauge field on the D7-branes is then, up to an overall normalisation, exactly the same as that considered in \cite{PSS} for a gauge field in $AdS_5$. The normalisation constant we obtain scales as $\tilde{\cal N}_\mt{D7} \sim \nf \nc$, whereas that of \cite{PSS} scales as $\nc^2$. This merely reflects the difference in the number of electrically charged degrees of freedom in the corresponding dual gauge theories. 

The solution of \eqn{A} satisfying the incoming-wave boundary condition was found in 
ref.~\cite{CKMSY}:
\be
A(\omega, u)=(1-2u)^{-i\wn/2}(1+2u)^{-\wn/2} \,_2 F_1 \left( a,b,c;  \frac{1-2u}{2} \right) \,, 
\label{solA}
\ee
where
\be
a = 1 - (1+i) \frac{\wn}{2} \sac b = -(1+i) \frac{\wn}{2} \sac c= 1-i \wn \,,
\ee
and $\,_2 F_1$ is a hypergeometric function. Substituting into eq.~\eqn{trace1} we obtain the result 
\be
\chi^\mu_{\,\,\mu} (\omega) = 
\frac{8 \tilde{\cal N}_\mt{D7} \, \wn}{2^\wn \left|_2 F_1(a,b,c;\frac{1}{2})  \right|^2} 
\label{top}
\ee
for the trace of the spectral density. A plot of the dimensionless combination 
$\chi^\mu_{\,\,\mu} (\omega) / 8 \tilde{\cal N}_\mt{D7} \, \wn$ is given by 
the top, solid, red curve in fig.~\ref{D3D7wplot}.
\FIGURE[h!]{
\includegraphics[width=    0.8 \textwidth]{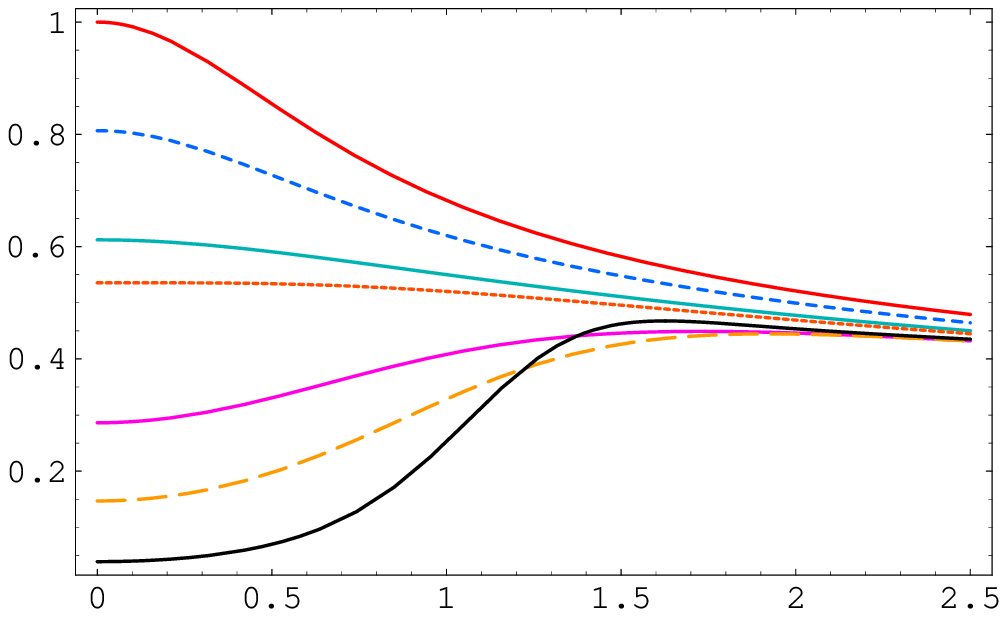}
\put(-180,-15){$\omega = k^0/ 2 \pi T$}
\put(-390,110){$\frac{\chi^\mu_{\,\,\mu}(\omega)}{8 \tilde{\cal N}_\mt{D7} \omega}$} 
\caption{D3/D7 system: Trace of the spectral function as a function of $\omega$ for (from top to bottom on the left-hand side)  $m=\{0, 0.6, 0.85, 0.93, 1.15, 1.25, 1.306 \}$, or equivalently for $\psi_0=\{0, 0.37, 0.53, 0.58, 0.75, 0.85, 0.941\}$. The last value corresponds to that at which the phase transition from a black hole to a Minkowski embedding takes place. Recall that $\tilde{\cal N}_\mt{D7}\sim  \nf \nc T^2$.}
\label{D3D7wplot} }
\FIGURE[h!!!]{
\includegraphics[width=    0.8 \textwidth]{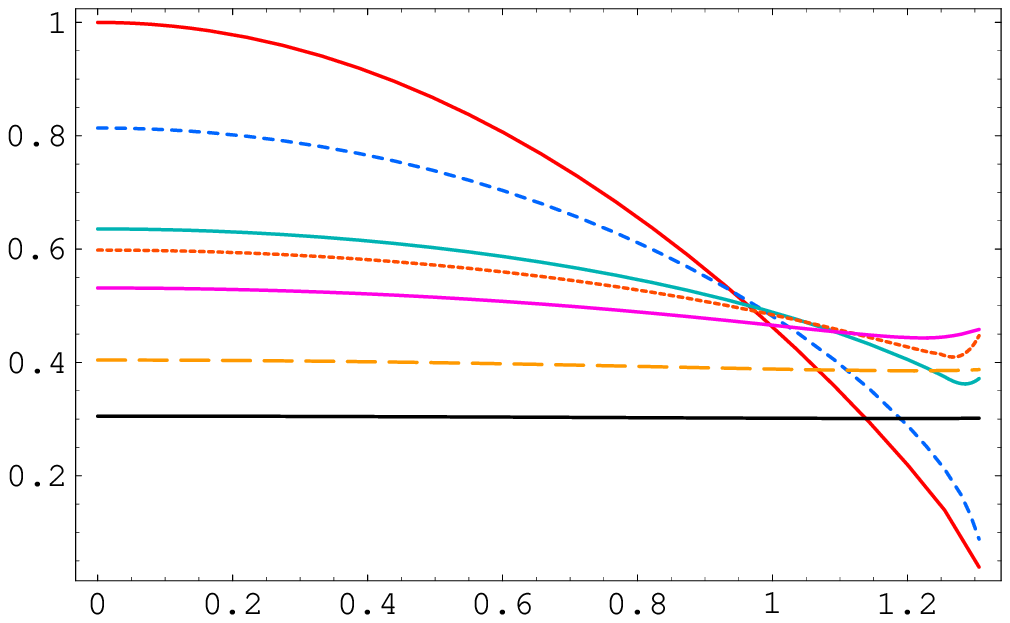}
\put(-180,-10){$m=\bar{M}/T$}
\put(-390,110){$\frac{\chi^\mu_{\,\,\mu}(\omega)}{8 \tilde{\cal N}_\mt{D7} \omega}$} 
\caption{D3/D7 system: Trace of the spectral function as a function of $m$ for (from top to bottom on the left-hand side) $\omega = \{0, 0.6, 1.2, 1.4, 1.9, 4, 9 \}$. The top, solid, red curve yields (up to normalisation) the electric conductivity.}
\label{D3D7mplot} }

\subsection{Massive quarks}
\label{D3D7massive}
The action \eqn{sdq} for the D7-branes in the absence of a gauge field takes the form
\be
S_\mt{D7} \propto  \int du \, \frac{1}{u^3} (1-\psi^2) \sqrt{1-\psi^2+4u^2f\psi'^2}  \,.
\label{sem}
\ee
By varying this with respect to $\psi(u)$ one obtains a non-linear, second-order differential equation for the D7-branes embedding. We were unable to solve this equation analytically for 
$\mq \neq 0$, but a numerical solution was obtained in \cite{prl,thermo}. This was done by integrating the differential equation from the horizon towards the boundary for different values of 
$\psi_0$. From the value of the solution at the boundary one then reads off $m(\psi_0)$ and $c(\psi_0)$ according to \eqn{asymp}. Using the solution $\psi(u)$ and the ansatz \eqn{sol} in \eqn{eom} a differential equation for $F$ is obtained, which can again be integrated numerically. Substituting the result in \eqn{trace1} yields the trace of the spectral function. Note that in 
eq.~\eqn{trace1} one should think of $\psi_0$ as a function of $m$.  

\FIGURE[t!!]{
\includegraphics[width=0.4 \textwidth]{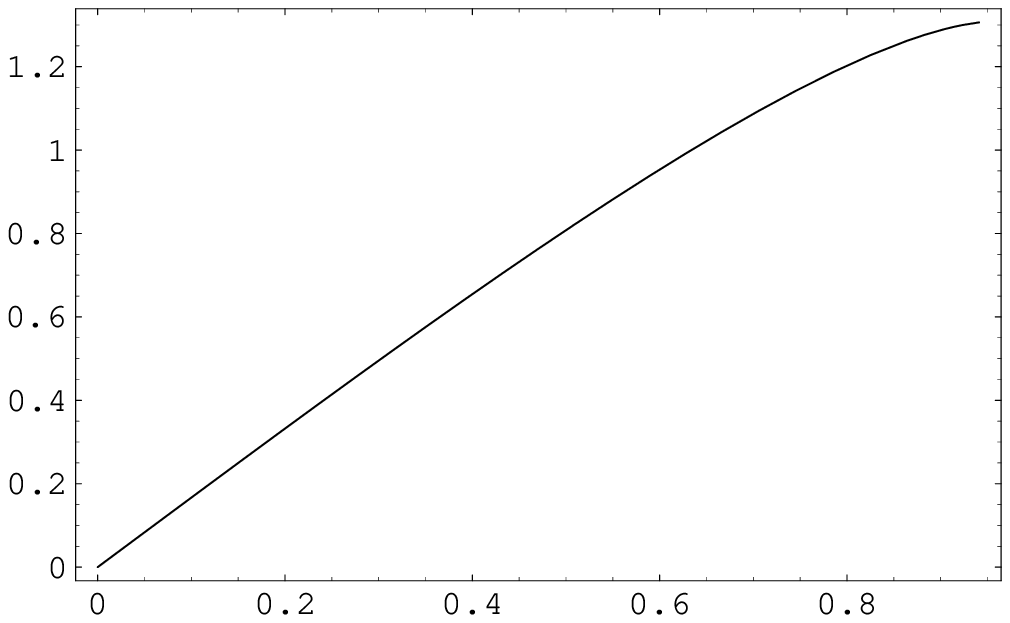}
\put(-85,-10){$\psi_0$}
\put(-190,50){$m$}
%\put(-220, 240){$m=0$}
\caption{Dimensionless ratio $m=\bar{M}/T$ as a function of $\psi_0$ in the range of values for which stable black hole embeddings exist.} 
\label{mBHplot}}
As shown in \cite{prl,thermo}, thermodynamically stable black hole embeddings for the D7-branes exist for $0 \leq \psi_0 < 0.941$, or equivalently for $0 \leq m < 1.306$, and in this parameter range $m$ is a monotonically increasing function of $\psi_0$, as exhibited in fig.~\ref{mBHplot}. 
For $\psi_0 > 0.941$ or $m > 1.306$ the free energy of the D7-branes is minimised by a Minkowski embedding, and so a first order phase transition occurs. Black hole embeddings with $0.941 < \psi_0 < 0.9621$, or equivalently with $1.306 < m < 1.3092$, are expected to be metastable \cite{melt,spectre}, whereas those with $\psi_0 > 0.9621$ have negative specific heat and are therefore unstable \cite{thermo}. 

Converting the differential emission rate \eqn{dgamma} to the emission rate per unit photon energy and using \eqn{trace1} gives
\be
\frac{d \Gamma}{d \, k^0} = 16 \aem \tilde{\cal N}_\mt{D7} T \, 
\frac{\left( 1-\psi_0^2 \right)^{3/2} w^2}{\left( e^{2\pi \omega} -1 \right)  
2^{\omega} \left| F(\omega , 0) \right|^2} \,.
\label{rate1}
\ee
Our results for the spectral function and for the photon emission rate for thermodynamically stable black hole embeddings are shown in figs.~\ref{D3D7wplot}, 
\ref{D3D7mplot}, and \ref{D3D7productionplot}. Some interesting features of these plots are as follows.
\FIGURE[h!]{
\begin{tabular}{c}
\includegraphics[width=    0.8 \textwidth]{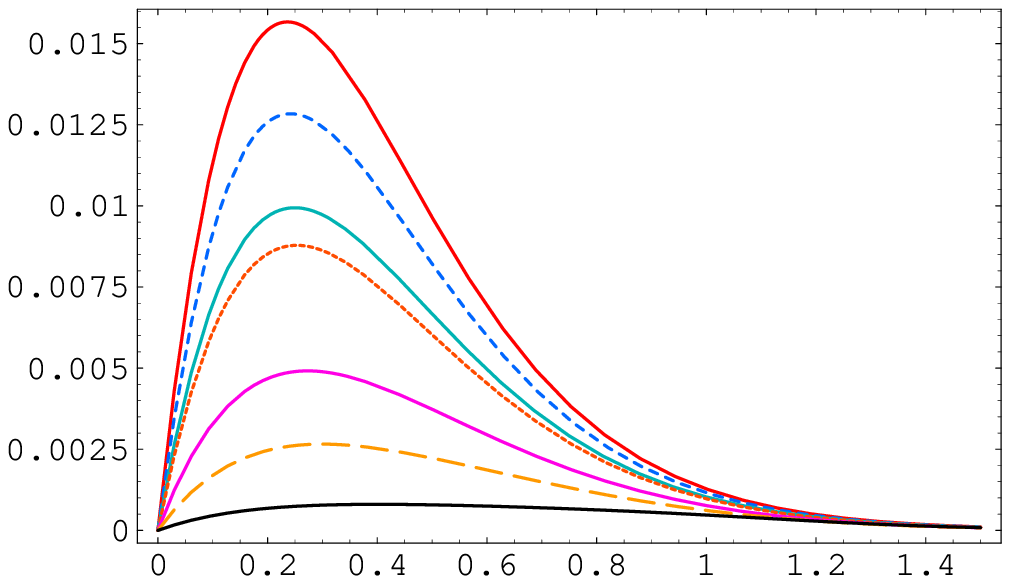}
\put(-180,-10){$\omega = k^0/ 2 \pi T$}
\put(-380,80){\rotatebox{90}{{$\frac{1}{16 \alpha_\mt{EM} \tilde{\cal N}_\mt{D7} T}
\frac{d \Gamma}{d \, k^0}$}}}
\vspace{5mm} \\
\includegraphics[width=    0.8 \textwidth]{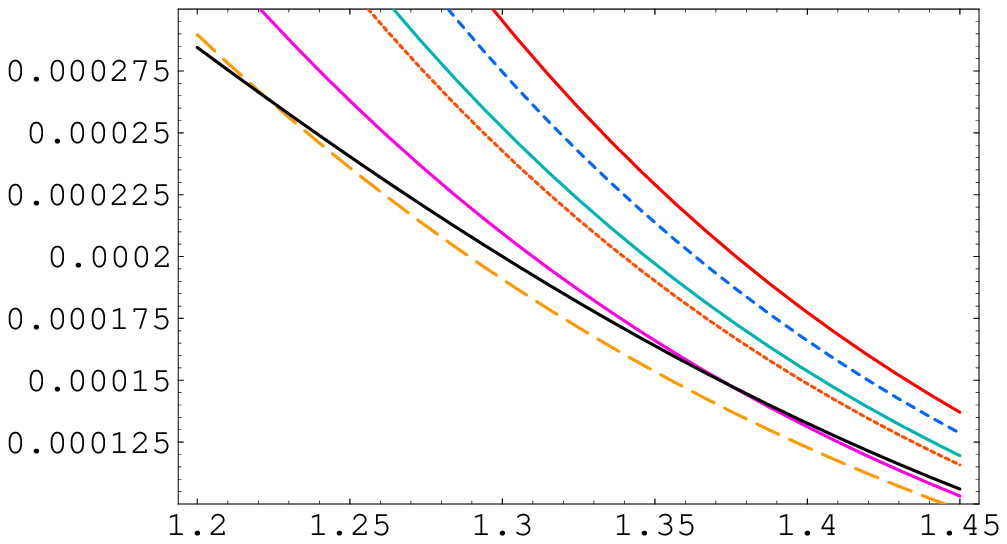}
\put(-180,-1){$\omega = k^0/ 2 \pi T$}
\put(-380,80){\rotatebox{90}{{$\frac{1}{16 \alpha_\mt{EM} \tilde{\cal N}_\mt{D7} T}
\frac{d \Gamma}{d \, k^0}$}}}
\end{tabular}
\caption{D3/D7 system: Photon emission rate as a function of $\omega$ for the same values of $m$ as in fig.~\ref{D3D7wplot}, \ie for (from top to bottom)  $m=\{0, 0.6, 0.85, 0.93, 1.15, 1.25, 1.306 \}$.}
\label{D3D7productionplot} } 

The top, solid, red curve in fig.~\ref{D3D7wplot} corresponds to the analytic result 
\eqn{top} for massless quarks. This is a monotonically decreasing function that for large frequencies decays as $\omega^{-1/3}$. All curves with $0 \leq m \lesssim 0.93$ are also monotonically decreasing functions of the frequency, whereas curves with 
$0.93 \lesssim m \leq 1.306$ are increasing functions of $\omega$ for small $\omega$. At large values of $\omega$ all curves seem to decay with the same power of the frequency as the $\mq=0$ curve, presumably due to the fact that at large $\omega$ the  scale associated to the quark mass becomes irrelevant. The behaviour at small 
$\omega$ can also be easily understood. As $\omega \rightarrow 0$ the suppression factor $\left( 1- \psi_0^2 \right)^{3/2}$ in eq.~\eqn{trace1} dominates, and so for small values of $\omega$ the spectral density decreases monotonically with $m$, as shown in fig.~\ref{D3D7mplot}. This effect is also responsible for the fact that the height of the peak of the photon emission curves in fig.~\ref{D3D7productionplot} largely decreases as $m$ increases from zero to its maximum value. The suppression factor above has a simple geometric origin as the decreasing area of the induced black hole horizon on the D7-branes as $\psi_0$ increases, as can be seen from eq.~\eqn{g7}. 

Note that the top, solid, red curve in fig.~\ref{D3D7mplot}, which corresponds to 
$\omega=0$, gives (up to normalisation) the electric conductivity \eqn{sigma}. Specifically, denoting by $h(m)$ the curve in question, one has:
\be
\sigma = \frac{e^2}{4(2\pi T)} \, \left. \frac{d\chi}{d \omega} \right|_{\omega=0} = 
\frac{e^2}{4 \pi} \nf \nc T \, h(m) \,.
\label{sigmaD3D7}
\ee
Again, the difference between our $\nf \nc$ scaling and the $\nc^2$ scaling found in 
\cite{CKMSY} reflects the difference in the number of electrically charged degrees of freedom. 

At intermediate values of $\omega$ the spectral function is not a monotonic function of $m$, as can be seen in fig.~\ref{D3D7mplot}. In fig.~\ref{D3D7wplot} this is reflected in the fact that curves for different values of $m$ cross each other around 
$1 \lesssim \omega \lesssim 2$. The same behaviour is of course observed in the plot of the photon production shown in fig.~\ref{D3D7productionplot}.

It is also interesting to examine the spectral function for black hole embeddings beyond the phase transition, \ie in the region in which these embeddings are metastable or unstable. The results for the spectral function are shown in fig.~\ref{D3D7wplotbis}. The most remarkable feature of these plots is the appearance of well defined peaks in the spectral function, which become narrower and more closely spaced, seemingly approaching delta-functions, as $\psi_0 \rightarrow 1$. We will discuss the interpretation of this fact in the last section. 
\FIGURE[h!]{
\includegraphics[width=    0.8 \textwidth]{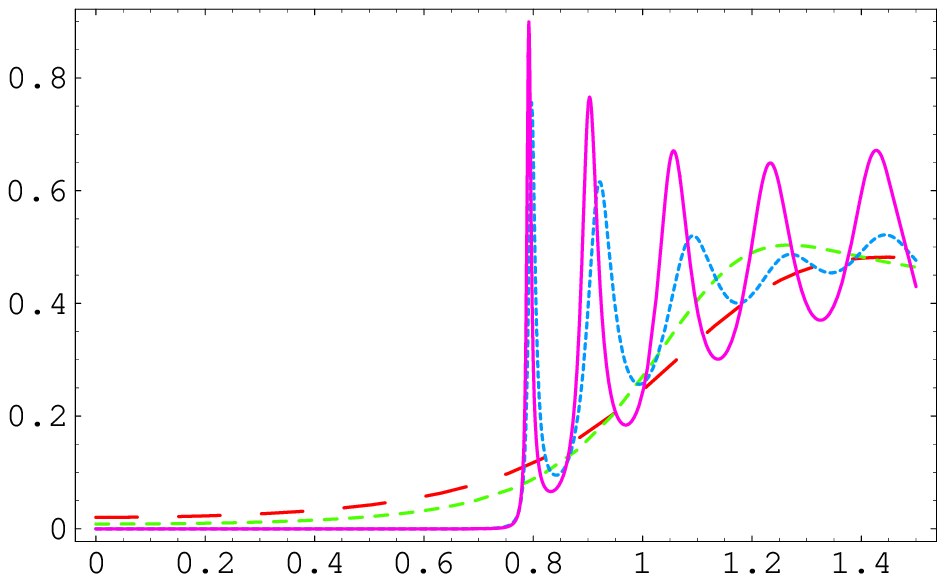}
\put(-180,-1){$\omega = k^0/ 2 \pi T$}
\put(-380,110){$\frac{\chi^\mu_{\,\,\mu}(\omega)}{8 \tilde{\cal N}_\mt{D7} \omega}$}
%\put(-220,-10){$\omega = k^0/ 2 \pi T$}
%\put(-460,130){$\frac{1}{16 \alpha_\mt{EM} \tilde{\cal N}_\mt{D7} T}
%\frac{d \Gamma}{d\omega}$}
\caption{D3/D7 system: Spectral function for non-stable black hole embeddings. 
The red curve with the longest dashes corresponds to $\psi_0=0.9621$, the green curve with intermediate dashes to $\psi_0= 0.979$, the blue curve with the shortest dashes to 
$\psi_0= 0.999996$, and the solid, purple curve to $\psi_0= 0.9999981$.}
\label{D3D7wplotbis} }

\section{The D4/D6 system}
\label{d4d6}

We now turn to the D4/D6 system, described by the array
\begin{equation}
\begin{array}{ccccccccccc}
   & 0 & 1 & 2 & 3 & 4& 5 & 6 & 7 & 8 & 9\\
D4 & \times & \times & \times & \times & \times & &  &  & & \\
D6 & \times & \times & \times & \times &   &\times &\times & \times &  &   \\
\end{array}\label{D4D6}
\end{equation}
In the decoupling limit, the resulting gauge theory is
five-dimensional SYM coupled to fundamental
hypermultiplets confined to a four-dimensional defect, so we have again $d=3$. 
In order to obtain a four-dimensional gauge theory at low energies, one may
compactify $x^4$, the D4-brane direction orthogonal to the defect,
on a circle. If periodic boundary conditions for the adjoint
fermions are imposed, then supersymmetry is preserved and the
four-dimensional theory thus obtained is non-confining. In this case
the appropriate dual gravitational background at any temperature is
\eqn{metric} with $x^4$ periodically identified. Instead, if
antiperiodic boundary conditions for the adjoint fermions are
imposed, then supersymmetry is broken and the four-dimensional
theory exhibits confinement \cite{witten} and spontaneous chiral
symmetry breaking \cite{us}. The holographic description at
zero-temperature consists then of D6-brane probes in a horizon-free
background, whose precise form is not needed here. At a temperature
$\tdec$ set by the radius of compactification, the theory undergoes a
first order phase transition at which the gluons and the adjoint
matter become deconfined. In the dual description the low-temperature background 
is replaced by \eqn{metric}. If $\tdec < \tf$, the D6-branes remain outside the horizon in a Minkowski embedding. As $T$ is further increased up to $\tf$ a first order phase transition
for the fundamental matter occurs.

The metric (\ref{metric}) for the black D4-brane is
\be
ds^2 = \frac{r^{3/2}}{L^{3/2}} \left( -f dx_0^2 +  d{\bf x}^2 \right) + 
\frac{L^{3/2}}{r^{3/2}} \frac{dr^2}{f} + r^{1/2}L^{3/2} d\Omega_{\it 4}^2 \,, 
\label{metricD4} 
\ee
where 
\be
L^3 = \pi g_s \nc \ell_s^3 \sac f= 1 -\frac{r_0^3}{r^3} \sac r_0={16\over 9}\pi^2 T^2 L^3 \,.
\ee
The five-dimensional Yang-Mills coupling constant in the dual gauge theory is dimensionful and given by $\gym^2 = 4\pi^2 g_s \ell_s$. It is convenient to introduce the dimensionless coordinate $u=r_0^{3/2}/2 r^{3/2}$, in terms of which the metric becomes
\be
ds^2 = \left(\frac{r_0}{L}\right)^{3/2} \frac{1}{2u} \left( -f dx_0^2 + d{\bf x}^2 \right) + 
\frac{L^{3/2} r_0^{1/2}}{\left( 2u \right)^{1/3}} \left( \frac{4 du^2}{9 u^2 f} 
+ d\Omega_{\it 4}^2 \right)  \,, 
\label{metricD4bis}
\ee
with $f=1-4u^2$. As before, the horizon is at $u =1/2$ and the boundary at 
$u \rightarrow 0$.\footnote{The D4-brane metric considered in this section is not asymptotically of the form $AdS$ times a sphere. The framework for the calculation of correlators is less well developed for such backgrounds, so we will proceed by analogy with the $AdS$ case. Presumably, however, this procedure can be made rigourous by lifting the D4-brane geometry to M-theory, in which it becomes an M5-brane geometry whose asymptotic form is $AdS_7 \times S^4$. The D6-brane lifts in turn to a KK-monopole, whose worldvolume effective action can be found in \cite{KK}.\label{caveat}} Since the D6-branes wrap a two-sphere in the directions transverse to the D4-branes, it is also useful to write the metric on the four-sphere as 
\be
d\Omega_{\it 4}^2 = d\theta^2 + \sin^2 \theta d\Omega_{\it 2}^2 + 
\cos^2 \theta d\varphi^2 \,,
\ee
and to set $\psi = \cos \theta$. The D6-brane embedding is then specified as $\psi=\psi(u)$, in terms of which the induced metric takes the form
\be
ds_\mt{D6}^2=\left( {4 \over 3} \pi TL\right)^3 \frac{1}{2u} \left( -fdx_0^2+ d{\bf x}^2 \right) + 
\frac{4 \pi T L^3}{3 \left( 2u \right)^{1/3}} 
\left[ \left( \frac{4 (1-\psi^2) + 9 u^2 f \psi'^2}{9 u^2 f (1-\psi^2)} \right) du^2 + 
(1-\psi^2) d\Omega_{\it 2}^2 \right] \,.
\label{metD6}
\ee
In eqs.~(\ref{metricD4}) and (\ref{metricD4bis}) ${\bf x}=(x^1, \ldots, x^4)$, whereas in \eqn{metD6} and all equations below in this section ${\bf x}=(x^1, \ldots, x^3)$.

As in the previous section, the branes embedding $\psi(u)$ is determined by extremising the action \eqn{sdq} with $F=0$. Note that in this case there is a non-constant dilaton, 
\be
e^{\phi} = \left(\frac{r_0}{L}\right)^{3/4} \frac{1}{\sqrt{2u}} \,,
\ee
reflecting the non-conformality of the dual gauge theory. Near the boundary,\footnote{In this limit, the coordinate $u$ used here and the coordinate $\rho$ used in 
\cite{prl,thermo} are related through $\rho^{3/2} = 1/u$.} $\psi(u)$ behaves as
\be
\psi(u) = m \, u^{2/3} + c \, u^{4/3} + \cdots \,,
\label{asympd6}
\ee
where the dimensionless constants $m$ and $c$ are proportional to the quark mass and condensate, respectively \cite{prl,thermo}:
\beqa 
\mq &=& \frac{r_0 m}{2^{5/3}\pi \ell_s^2}=
{2^{1/3}\over3^2}\,\leff(T)^2\,T\,m \,, \label{mc4} \\
\langle {\cal O} \rangle &=& -2^{5/3}\pi^2 \ell_s^2 \nf T_\mt{D6}
r_0^2 c = -\frac{2^{5/3}}{3^4}\,\nf\,\nc\,\leff(T)^2 T^3 c \,,\label{donc4}
\eeqa
with
\be
\leff(T)^2 = \lambda T = \gym^2 \nc T
\ee
the effective 't Hooft coupling at the scale $T$. In this case, we may write 
$m=\mbar^2/T^2$ with
\be 
\mbar^2 = \frac{9}{2^{1/3} }
\left(\frac{\mq}{\leff(\mq)}\right)^2 \simeq
7.143\left(\frac{\mq}{\leff(\mq)}\right)^2\,. 
\ee
The scale $\mbar$ is again related to the mass gap in the meson
spectrum of the D4/D6 system at zero temperature through
$\mbar \simeq 0.233 \, \mmes$ \cite{holomeson2}.

Following the previous section one obtains the equation of motion \eqn{eom} with 
the functions $P(u), Q(u)$ given in this case by 
\be
P(u) = \frac{\sqrt{1-\psi^2} \sqrt{1-\psi^2 + {9\over 4}u^2 f \psi'^2}}{u f (2 \pi T)^2}  \sac
Q(u) = \frac{(2u)^{1/3} (1-\psi^2)^{3/2} f}{2 \sqrt{1-\psi^2 + {9\over 4}u^2 f \psi'^2}} \,.
\ee
This equation of motion may be derived from the action \eqn{d7s}, where
\be
\tilde{\cal N}_\mt{D6}= {\cal N}_\mt{D6} \frac{3T (2\pi\ell_s^2)^2}
{\left( \frac{16}{9} \pi^2 T^2 L^3 \right)^2} = \frac{1}{3} \nf \nc T^2 \,, 
\ee
and 
\be
{\cal N}_\mt{D6} = \frac{2^2}{3^6} \nf \nc \leff(T)^4 T^3
\ee
is the normalisation constant introduced in \cite{prl,thermo}. In this case 
eq.~\eqn{lim} becomes
\be
\lim_{u \rightarrow \frac{1}{2}}  2Q(u) A^*(\omega, u) \partial_u A(\omega, u) = 
\frac{2 i \wn \left( 1- \psi_0^2 \right)}{2^\wn} 
\ee
and the trace of the spectral density, eq.~\eqn{trace}, takes the form
\be
\chi^\mu_{\,\,\mu} (\omega) = -4 \, \mbox{Im} \, G^\mt{R}(\omega) = 
8 \tilde{\cal N}_\mt{D6}  \left( 1- \psi_0^2 \right) \,
\frac{\wn}{2^\wn \left| F(\omega,0) \right|^2} \,.
\label{trace2}
\ee

The action \eqn{sdq} for the D6-branes in the absence of a gauge field reduces to 
\be
S_\mt{D6} \propto  \int 
du \, \frac{1}{u^3} \sqrt{1-\psi^2} \sqrt{1-\psi^2+ \frac{9}{4} u^2 f \psi'^2}  \,.
\label{sembis}
\ee
From this point one proceeds in complete analogy with the D3/D7 system, \ie one solves for the D6-brane embedding numerically and uses the result to solve, also numerically, for the function $F(\omega, u)$. Substituting this in \eqn{trace2} one obtains the trace of the spectral density. 

In this case the phase transition takes place at $\psi_0=0.822$ or $m=1.589$. Below these values black hole embeddings are stable, whereas for $0.822 < \psi_0 < 0.907$, or $1.589 < m < 1.630$, they are presumably metastable. For $\psi_0 > 0.907$ black hole embeddings possess a negative specific heat and are therefore unstable \cite{thermo}.
The results for stable embeddings are plotted in figs.~\ref{D4D6wplot},
\ref{D4D6mplot} and \ref{D4D6productionplot}, whereas those for metastable or unstable embeddings are shown in fig.~\ref{D4D6wplotbis}. Remarkably, these plots share many of the qualitative features of the D3/D7 system. We will come back to this in the next section.
\FIGURE[t!]{
\includegraphics[width=    0.8 \textwidth]{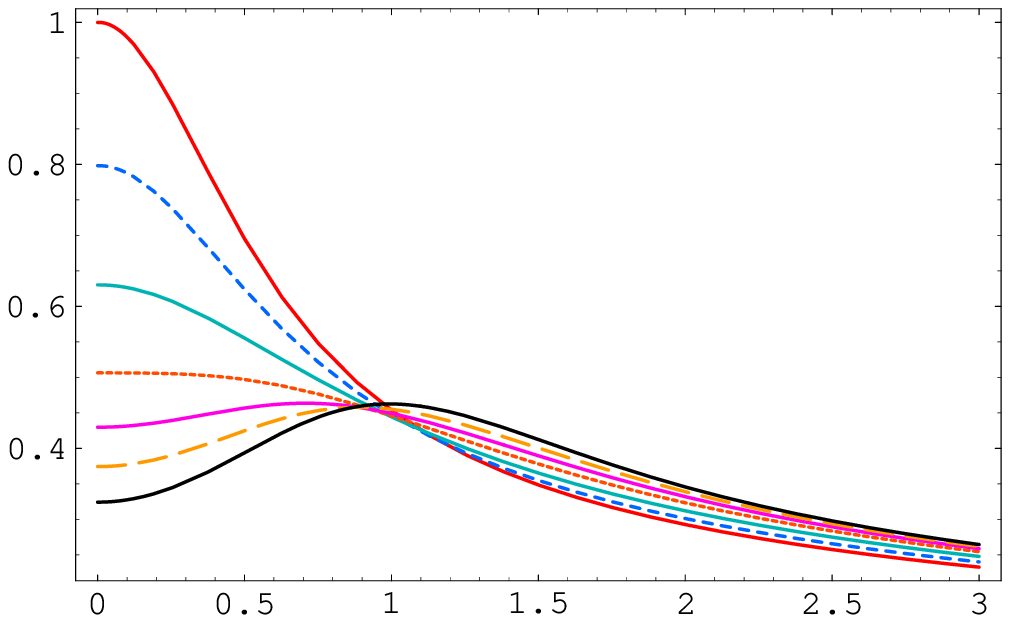}
\put(-180,-15){$\omega = k^0/ 2 \pi T$}
\put(-390,110){$\frac{\chi^\mu_{\,\,\mu}(\omega)}{8 \tilde{\cal N}_\mt{D6} \omega}$} 
\caption{D4/D6 system: Trace of the spectral function as a function of $\omega$ for (from top to bottom on the left-hand side)  $m=\{0, 1, 1.3, 1.45, 1.52, 1.56, 1.589 \}$, or equivalently for $\psi_0=\{0, 0.449, 0.608, 0.703, 0.755, 0.791, 0.822\}$. }
\label{D4D6wplot} }
\FIGURE[h!]{
\includegraphics[width=    0.8 \textwidth]{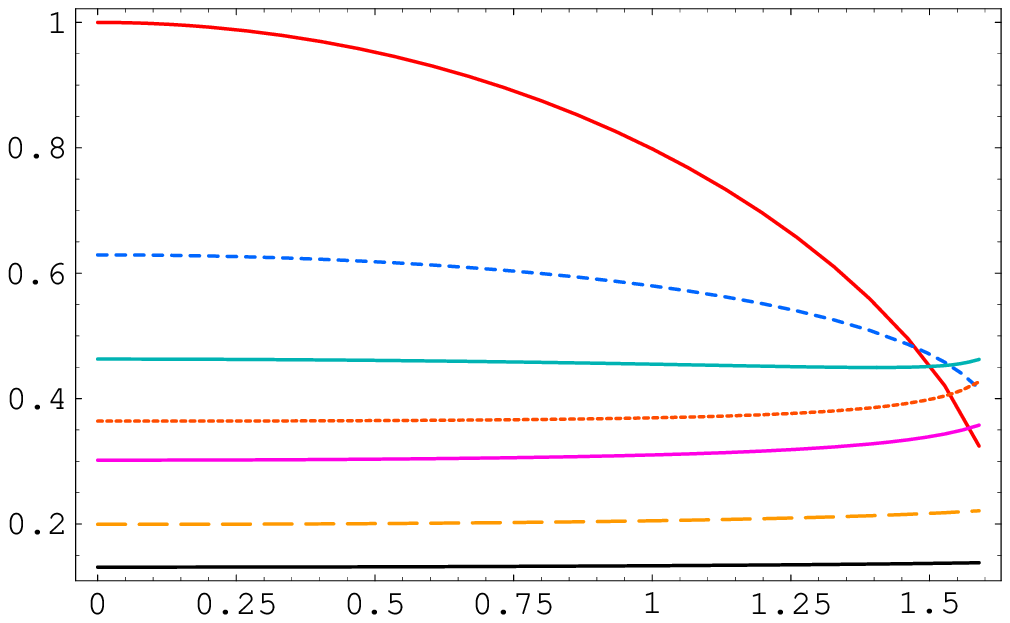}
\put(-180,-15){$m=\bar{M}^2/T^2$}
\put(-390,110){$\frac{\chi^\mu_{\,\,\mu}(\omega)}{8 \tilde{\cal N}_\mt{D6} \omega}$} 
\caption{D4/D6 system: Trace of the spectral function as a function of $m$ for (from top to bottom on the left-hand side) $\omega = \{0, 0.6, 0.97, 1.4, 1.9, 4, 9 \}$. The top, solid, red curve yields (up to normalisation) the electric conductivity.}
\label{D4D6mplot} }

\FIGURE[h!]{
\begin{tabular}{c}
\includegraphics[width=    0.8 \textwidth]{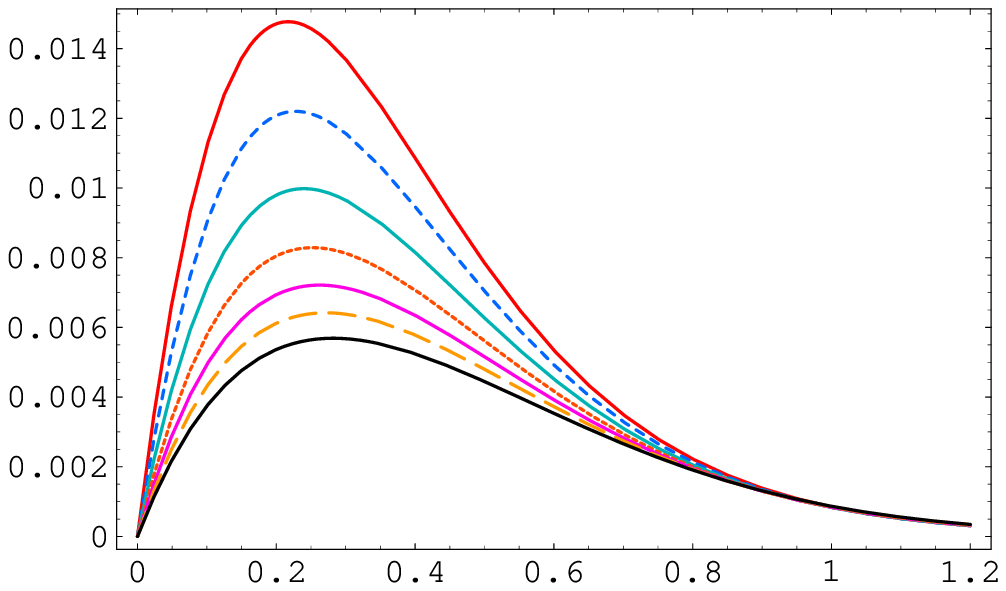}
\put(-180,-10){$\omega = k^0/ 2 \pi T$}
\put(-380,80){\rotatebox{90}{{$\frac{1}{16 \alpha_\mt{EM} \tilde{\cal N}_\mt{D6} T}
\frac{d \Gamma}{d \, k^0}$}}}
\vspace{5mm} \\
\includegraphics[width=    0.8 \textwidth]{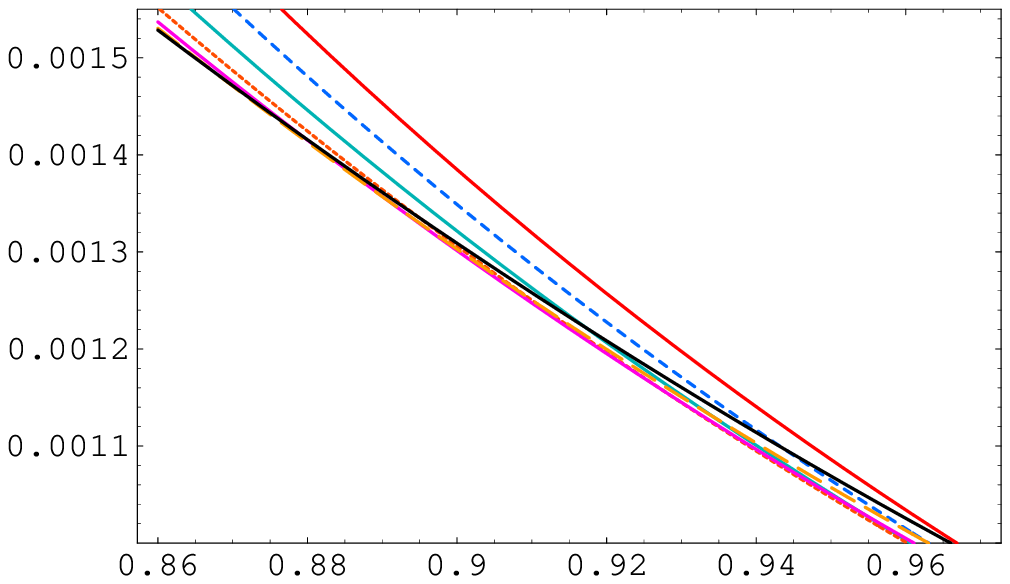}
\put(-180,-10){$\omega = k^0/ 2 \pi T$}
\put(-380,80){\rotatebox{90}{{$\frac{1}{16 \alpha_\mt{EM} \tilde{\cal N}_\mt{D6} T}
\frac{d \Gamma}{d \, k^0}$}}}
\end{tabular}
\caption{D4/D6 system: Photon emission rate as a function of $\omega$ for the same values of $m$ as in fig.~\ref{D4D6wplot}, \ie for (from top to bottom)  $m=\{0, 1, 1.3, 1.45, 1.52, 1.56, 1.589 \}$.}
\label{D4D6productionplot} }

\FIGURE[t!]{
\includegraphics[width=    0.8 \textwidth]{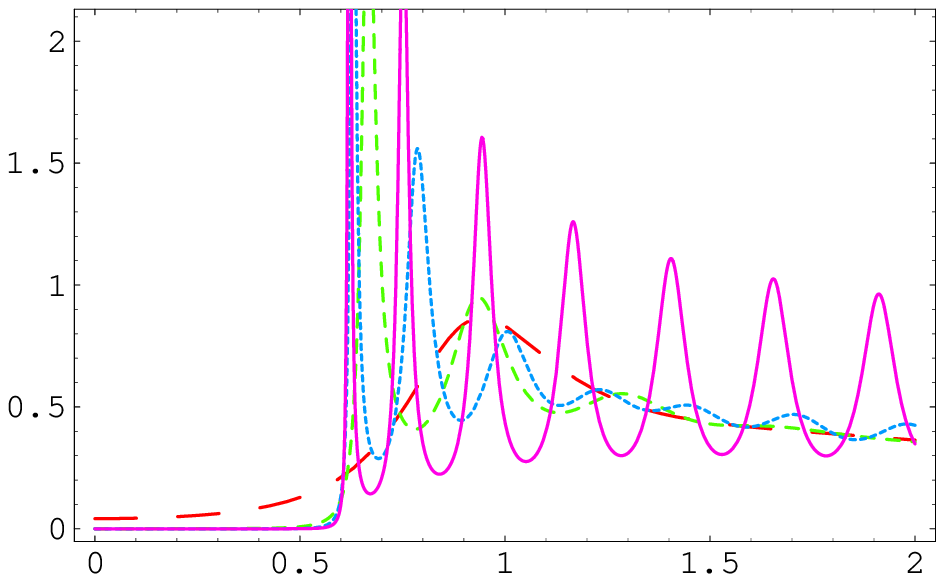}
\put(-180,-1){$\omega = k^0/ 2 \pi T$}
\put(-380,120){$\frac{\chi^\mu_{\,\,\mu}(\omega)}{8 \tilde{\cal N}_\mt{D6} \omega}$} 
\caption{D4/D6 system: Spectral function for non-stable black hole embeddings. 
The red curve with the longest dashes corresponds to $\psi_0=0.979$, the green curve with intermediate dashes to $\psi_0= 0.9999$, the blue curve with the shortest dashes to 
$\psi_0= 0.999996$, and the solid, purple curve to $\psi_0= 0.9999993$. }
\label{D4D6wplotbis} }

\section{Discussion}
\label{discussion}

Results from heavy ion collision experiments at RHIC indicate that the QGP created there behaves as a strongly coupled liquid. In the absence of systematic, non-perturbative methods to calculate real-time observables in QCD, it is useful to calculate these observables in gauge theories for which a gravity dual is known, the hope being that universal properties of such theories may be shared by QCD itself. 

In this paper we have calculated retarded correlators of electromagnetic currents at light-like momenta in a class of large-$\nc$, strongly coupled $SU(\nc)$ SYM theories with 
$\nf$ flavours of fundamental matter. From these correlators one may extract several observables associated to photon production. Our results are valid to leading order in the electromagnetic coupling constant and in $\nf/\nc$, but non-perturbatively in the 
$SU(\nc)$ interactions. At a qualitative level, these results exhibit a number of universal features. 

For small values of the quark mass the combination $\chi^\mu_{\,\,\mu} (\omega)/ \omega$ is a featureless, monotonically decreasing function of $\omega$, as observed in figs.~\ref{D3D7wplot} and \ref{D4D6wplot}. Instead, for large enough values of $m$, 
$\chi^\mu_{\,\,\mu} (\omega) / \omega$ first increases with $\omega$, then reaches a maximum and finally decays  as $\omega^{-\delta}$ for large $\omega$. The exponent is however not universal: For the D3/D7 system one has $\delta = 1/3$, whereas for the 
D4/D6 one finds $\delta = 2/3$. 

The qualitative behaviour of $\chi^\mu_{\,\,\mu} (\omega) / \omega$ as a function of $m$ for fixed values of $\omega$ also exhibits universal features. For small $\omega$ this is a monotonically decreasing function of $m$, as displayed in figs.~\ref{D3D7mplot} and 
\ref{D4D6mplot}; in particular, the top, solid, red curves in these figures, which correspond to $m=0$, yield (up to normalisation) the electric conductivity.\footnote{In the D4/D6 system the electric conductivity is $\sigma = (e^2/3\pi) \nf \nc T \, h(m)$, where 
$h(m)$ is the top, solid, red curve in fig.~\ref{D4D6mplot}.} Instead, for intermediate values of $\omega$, the spectral function as a function of $m$ ceases to be monotonic. In figs.~\ref{D3D7wplot} and \ref{D4D6wplot} this is reflected in the fact that curves for different values of $m$ cross each other. This is also the case for the photon production curves, as can be seen in figs.~\ref{D3D7productionplot} and \ref{D4D6productionplot}. This is perhaps a counterintuitive result, as it means that, at certain frequencies, plasmas with heavier quarks glow more brightly than those with lighter quarks. 

Above we have calculated the rate of photon emission of an homogeneous, infinitely extended, thermally equilibrated plasma. In order to extract a prediction for a physical situation such as that of a heavy ion collision experiment, our results would have to be integrated over the spacetime evolution of the plasma created in such experiments, 
as determined by some hydrodynamic model \cite{evolution}. In such a situation the temperature changes over time (and space) from some initial, maximum temperature $T_\mt{max}$, to some final, minimum temperature $T_\mt{min}$, and one measures the total number of produced photons of a given energy. This means that, unlike in the plots presented above, both $\omega$ and $m$ vary simultaneously as 
$\omega \propto T^{-1}$ and $m \propto T^{-1}$ (for the D3/D7 system) or 
$m \propto T^{-2}$ (for the D4/D6 system). Temperatures around 
$T_\mt{max} \simeq 4 \tdec$ are expected to be achieved at LHC, whereas those at RHIC are roughly around $2 \tdec$, where $\tdec \simeq 175$ MeV is the deconfinement temperature in QCD. On the other hand, $\tdec$ is the lowest temperature for which a plasma description is appropriate, since below such a temperature hadronisation occurs. 
For illustrative purposes, let us therefore consider the photon emission predicted by our models as a function of $T$, for $\tdec \leq T\leq 4 \tdec$ and for fixed values of $k^0$ and $\mq$. It is then convenient to write
\be
\omega = \gamma_1 x \sac \gamma_1 = \frac{k^0}{2\pi \tdec}  \sac 
x = \frac{\tdec}{T} \,.
\ee
Additionally, for the D3/D7 system we have
\be
m = \gamma_2 x \sac \gamma_2 = \frac{\mmes}{2 \pi \tdec} \,,
\ee
whereas for the D4/D6 system
\be
m = \gamma_2 x^2 \sac \gamma_2 = \frac{0.233^2 \mmes^2}{\tdec^2} \,.
\ee
Recall that $\mmes \propto \mq$ is the typical mesonic scale in the gauge theory at zero temperature. 

We are interested in the rate of photon production as $x$ varies between 1/4 and 1 for fixed values of $\gamma_1, \gamma_2$. Converting the differential emission rate 
\eqn{dgamma} to the emission rate per unit photon energy and using the results 
\eqn{trace1}, \eqn{trace2} for the spectral functions gives
\be
\frac{ \beta_1 \pi^2}{ \aem \nf \nc (k^0)^2 \tdec} \,
\frac{d \Gamma}{d \, k^0} = 
\frac{(1-\psi_0^2)^{\beta_2}}{x \left( e^{2\pi \gamma_1 x} -1 \right) 
2^{\gamma_1 x} \left| F(\gamma_1 x , 0) \right|^2} \,,
\label{rate}
\ee
where $\{ \beta_1 =1, \beta_2 = 3/2 \}$ for the D3/D7 system, 
$\{ \beta_1 =3/4, \beta_2 = 1 \}$ for the D4/D6 system, and the factor $(1-\psi_0^2)$ must be understood as a function of $m(x)$. 

The right-hand side of eq.~\eqn{rate} for the D3/D7 system is plotted in 
fig.~\ref{D3D7realplot} for several values of the photon energy and the mesonic scale 
$\mmes$. Specifically, the top graph corresponds to $k^0=100$ MeV, whereas the bottom graph corresponds to $k^0 = 1000$ MeV. The meson masses correspond to representative values in QCD for bound states of $u$ and $d$ quarks, $s$ quarks and 
$c$ quarks: $\mmes = M_\pi (140 \mbox{ MeV}), M_\phi (1020 \mbox{ MeV}), 
M_{J/\psi} (3096 \mbox{ MeV})$. For comparison, we have also included curves for 
$\mmes =0$ and $\mmes = M^*$, where $M^*\simeq 0.766 \tdec \simeq 1435$ MeV is the critical mass (corresponding to $m=1.306$) for which the phase transition to a Minkowski embedding takes place exactly at $T=\tdec$. For $\mmes =  M_{J/\psi}$ this transition takes place at $\tf \simeq 377 \mbox{ MeV} > \tdec$, and this is the reason why the dotted, orange curve terminates at $x =175 / 377 \simeq 0.464$: At this temperature the corresponding quarks form mesonic bound states, \ie in the string description the probe D7-branes jump from a black hole to a Minkowski embedding. 
\FIGURE[t!]{
\begin{tabular}{c}
\includegraphics[width=    0.8 \textwidth]{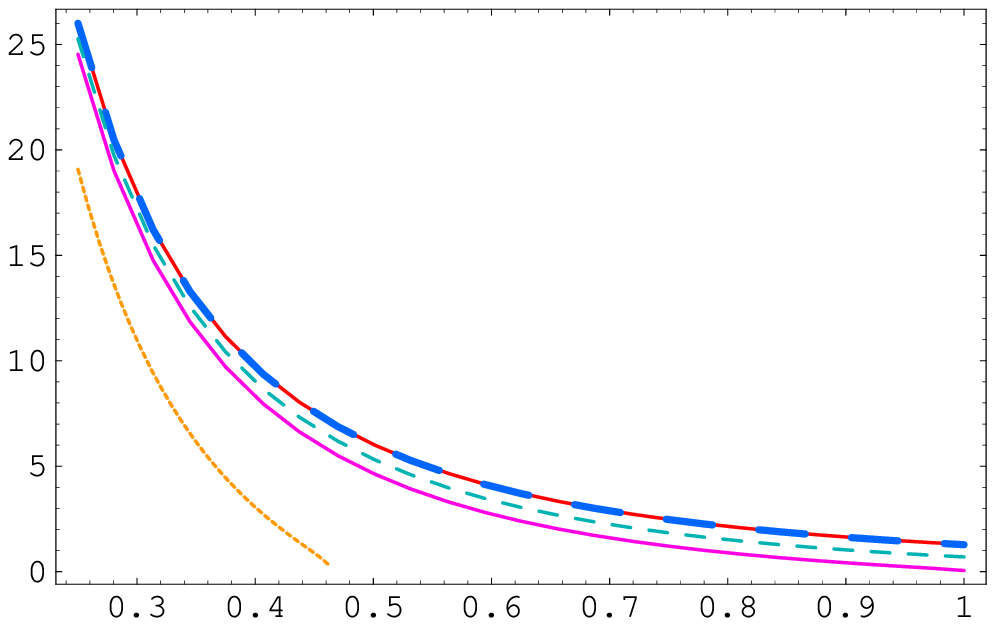}
\put(-180,-15){$x=\tdec / T$}
\put(-85,195){$k^0 = 100$ MeV}
\put(-380,60){\rotatebox{90}{{$\frac{\pi^2}{\alpha_\mt{EM} \nf \nc (k^0)^2 \tdec} 
\frac{d\Gamma}{d \, k^0} $}}}
\vspace{5mm} \\
\includegraphics[width=    0.8 \textwidth]{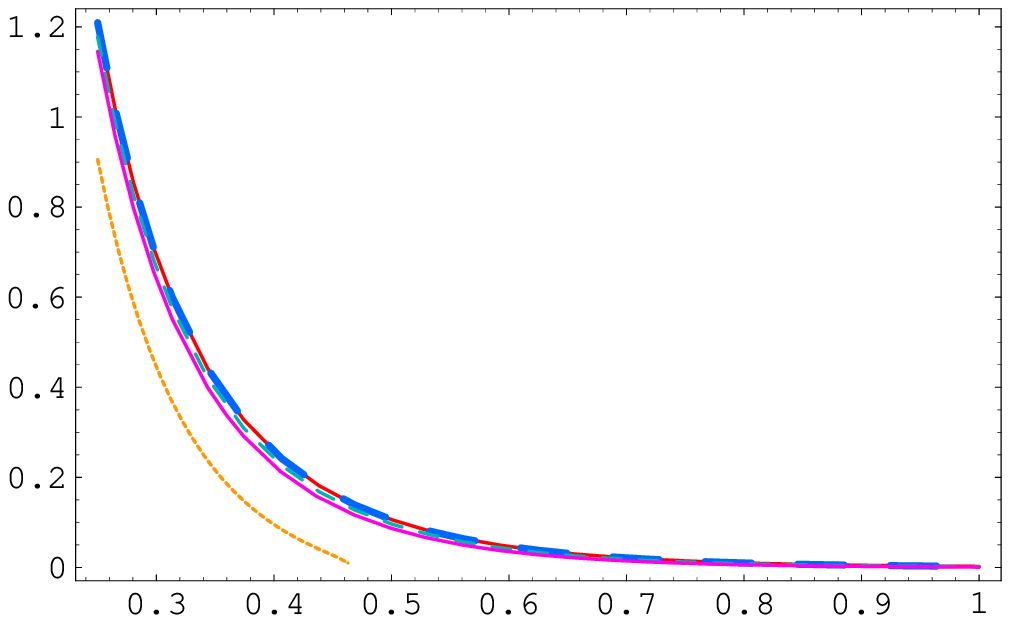}
\put(-180,-15){$x=\tdec / T$}
\put(-88,195){$k^0 = 1000$ MeV}
\put(-380,60){\rotatebox{90}{{$\frac{\pi^2}{\alpha_\mt{EM} \nf \nc (k^0)^2 \tdec} 
\frac{d\Gamma}{d \, k^0} $}}}
\caption{D3/D7 system: Photon emission rate as a function of $T$ for fixed $k^0$ and 
$\mq$.}
\end{tabular}
\caption{D3/D7 system: Photon emission rate as a function of $T$ for fixed $k^0$ and 
$\mmes$. The top graph corresponds to $k^0=100$ MeV, whereas the bottom one corresponds to $k^0 = 1000$ MeV. From top to bottom, the different curves correspond to $\mmes = 0, 140, 1020,  1435, 3096$ MeV. Note that the first two curves (solid red and blue dashed) are virtually coincident.}
\label{D3D7realplot} }
Fig.~\ref{D4D6realplot100} displays an analogous plot for the D4/D6 system, for $k^0=100$ MeV. In this case the dashed, green line also terminates at $T \simeq 188 \mbox{ MeV} > \tdec$, since for the D4/D6 system both the $J/\psi$ and the $\phi$ mesons survive as bound states above the deconfinement temperature \cite{thermo}. 
\FIGURE[t!]{
\includegraphics[width=    0.8 \textwidth]{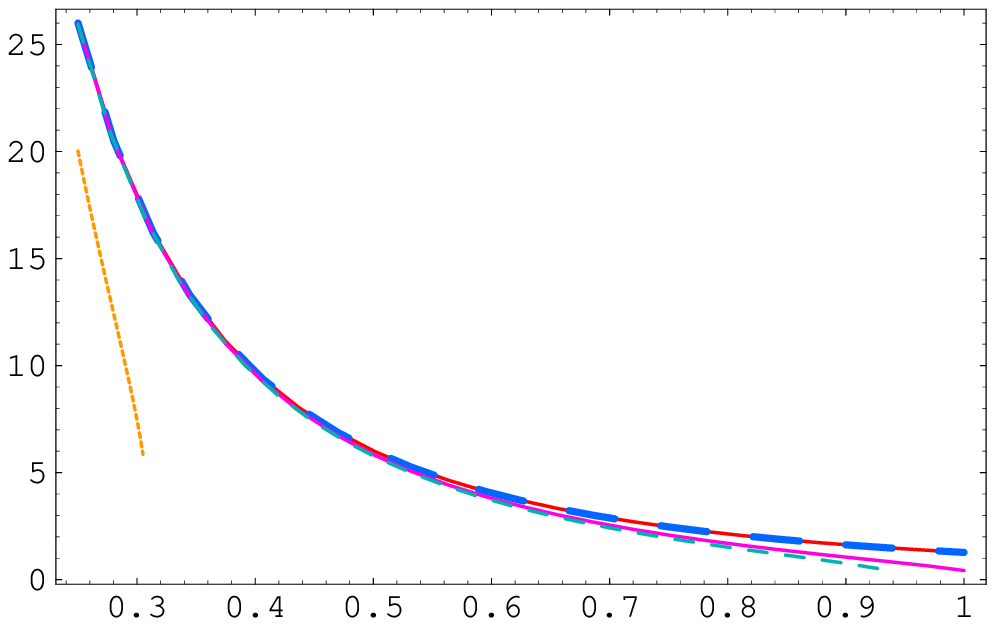}
\put(-180,-15){$\tdec / T$}
\put(-85,195){$k^0 = 100$ MeV}
\put(-380,60){\rotatebox{90}{{$\frac{3 \pi^2}{4 \alpha_\mt{EM} \nf \nc (k^0)^2 \tdec} 
\frac{d\Gamma}{d \, k^0} $}}}
\caption{D4/D6 system: Photon emission rate as a function of $T$ for $k^0=100$ MeV and several values of $\mmes$: From top to bottom, the different curves correspond to 
$\mmes = 0, 140, 947, 1020, 3096$ MeV. Note that the first two curves (solid red and blue dashed) are virtually coincident.}
\label{D4D6realplot100} }

One common feature of both models is that the rate of photon production is most sensitive to the quark mass for low-energy photons and low temperatures. As we see in fig.~\ref{D3D7realplot}, the difference between the curves is greater for $k^0=100$ MeV than for $k^0=1000$ MeV, and in the former case it is greater for low temperatures. 
We have verified that as $k^0$ increases above 1000 MeV, the difference between curves becomes smaller and smaller. The same conclusion holds for the D4/D6 system.
It is also remarkable that, contrary to what one might have perhaps expected, 
the rate of photon production associated to heavier quarks is smaller than that of lighter quarks, but it is not negligibly smaller. For example, we see in the figs. above that the number of photons emitted by $c$ quarks is a significant fraction of the number of photons emitted by $u$ and $d$ quarks for a wide range of temperatures. 

Before leaving our discussion of possible implications for heavy ion collision experiments, we would like to emphasise that our results apply to photons produced by a thermally equilibrated plasma. However, in heavy ion collisions there are other sources of photon production. For example, prompt photons are produced by initial collisions of the partons that constitute the colliding ions, whereas decay photons are produced by the decay of certain hadrons once the plasma ball has hadronised. Thus comparison of any theoretical prediction for thermal photons to empirical data requires being able to distinguish experimentally between these different sources. 

The features described above refer to thermodynamically stable black hole embeddings, but we also obtained results for metastable and unstable embeddings. These embeddings are skipped over by the meson-melting phase transition, but they are still interesting since they illustrate how the spectral function approaches that of Minkowski embeddings, as discussed in  \cite{spectre}. Results for the spectral function for several near-critical embeddings, \ie for values of $\psi_0$ close to 1, are displayed in figs.~\ref{D3D7wplotbis} and \ref{D4D6wplotbis}. Clearly the spectral function 
develops closely spaced, narrow peaks that seemingly approach delta-functions. 
Thus the form of the spectral function appears to approach the form we expect for Minkowski embeddings,\footnote{An analogous result was found in \cite{spectre} for time-like momenta.} namely an infinite sum of delta functions supported at a discrete set of energies $k^0= |{\bf k}|$. Each of these delta-functions is associated to a meson mode on the Dq-branes with null momentum. The existence of these modes may seem surprising in view of the fact that, as expected on general grounds and as verified in 
ref.~\cite{thermo}, the meson spectrum in the Minkowski phase possesses a mass gap, but in fact it follows from the analysis in ref.~\cite{thermo}. To see this, consider the dispersion relation $k^0 (\bk)$ for a given meson in the Minkowski phase. The fact that there is a mass gap means that $k^0 > 0$ at $\bk =0$. On the other hand, in the limit of infinite spatial momentum, $|{\bf k}| \rightarrow \infty$, the dispersion relation takes the form $k^0 \simeq v |{\bf k}|$ with $v < 1$. The reason for this is easily understood: For larger and larger spatial momenta, the wave function of the meson becomes more and more concentrated at the tip of the Dq-branes, and so the speed of the meson is simply the local speed of light at this lowest point. Because of the gravitational redshift, this speed is always subluminal. Continuity then implies that there must exist a value of 
$\bk$ such that $k^0 (|\bk|) = |\bk|$. 
\FIGURE[t!]{
\includegraphics[width=    0.8 \textwidth]{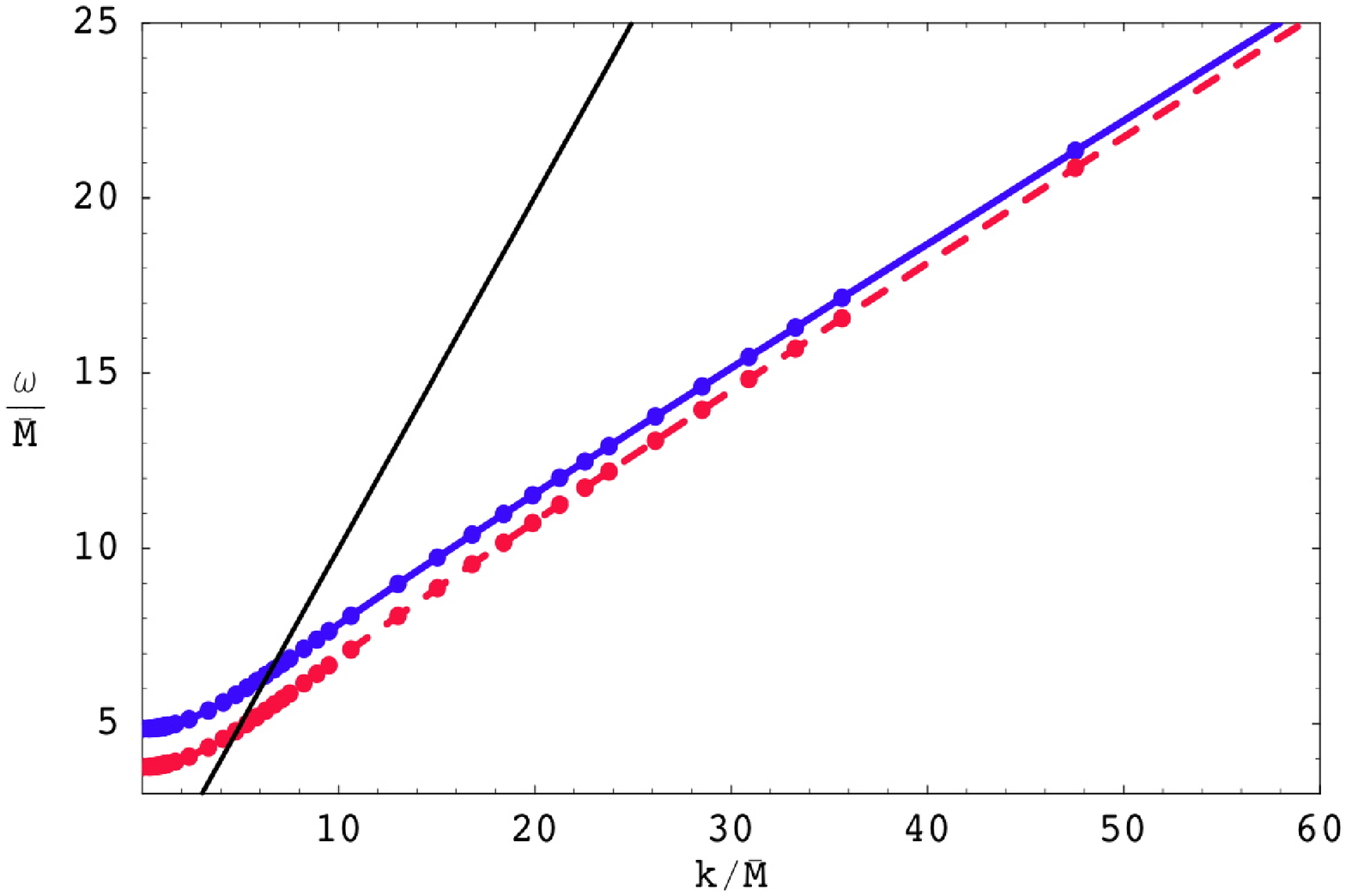}
\caption{Dispersion relation $\omega(|k|)$ for the lightest spin-zero mesons on  a
D7-brane Minkowski embedding in a D3-brane background \cite{thermo}. The solid blue curve corresponds to a pseudo-scalar meson, whereas the red dashed curve corresponds to a scalar meson. The solid black line corresponds to $\omega = |\bk|$.}
\label{dispersion-relation} }
This is illustrated by the fact that the solid black line in fig.~\ref{dispersion-relation} intersects the other two curves. Although the curves shown in the fig.~\ref{dispersion-relation} correspond to scalar mesons, it is clear from the arguments above that an analogous result would hold for other types of mesons, in particular for vector-like mesons. Since these mesons are absolutely stable in 
the large-$\nc$, strong coupling limit under consideration, they give rise to delta-function-like (\ie zero-width) peaks in the spectral function of electromagnetic currents.  

In this paper we have studied photon production by calculating the electromagnetic current-current correlator \eqn{green} at light-like momenta. It would be interesting to extend the calculation to time-like momenta, since this determines the nuumber of dileptons produced by virtual photon decay. A first step in this direction was given in 
ref.~\cite{spectre}, where the correlator was calculated for vanishing spatial momentum. 

A further extension consists of calculating the rate of photon and dilepton production in the presence of a finite baryon chemical potential \cite{potential-others,potential} of density \cite{density,density-others}. For non-zero baryon density black hole embeddings exist for all values of $T$ and $\mq$, and so for large $\mq / T$ the spectral function of electromagnetic currents should again reveal high and narrow peaks corresponding to the existence in the spectrum of long-lived vector mesons \cite{potential}.

\acknowledgments 
It is a pleasure to thank S.~Hartnoll, P.~Kovtun, R.~Myers, D.T.~Son, A.~Starinets and L.~Yaffe for helpful discussions. We are also grateful to R.~Myers for a careful reading of this manuscript, and to the authors of \cite{spectre} for sharing their work with us prior to publication. DM acknowledges support from NSF grant No PHY-0555669. 
LP acknowledges support from NSF grant No PHY03-54978 and from PROFIP (UNAM, Mexico).

\end{document}